\documentclass[12pt]{article}

\usepackage{amssymb}%,natbib}
\usepackage{graphicx}
\usepackage{amsmath}
\usepackage{amsfonts,amsthm}
\usepackage{ulem}

\graphicspath{{fig/}}

\newcommand{\tr}{\mathop{\rm tr}}

\newtheorem{theorem}{Theorem}
\newtheorem{corollary}{Corollary}
\newtheorem{lemma}{Lemma}
\newtheorem{definition}{Definition}

\newtheorem{remark}{Remark}

\usepackage{color}

\def\vec{{\rm vec}}
\def\cov{{\rm cov}}

\def\hat{\widehat}

\def\th{{\rm th}}
\def\rmspan{{\rm span}}

\def\tr{{\rm tr}}
\def\IF{{\rm IF}}

\title{Robust self-tuning semiparametric PCA for contaminated elliptical distribution}
\author{}
\date{\today}

\begin{document}
\author{Hung Hung$^{1}$, Su-Yun Huang$^{2}$
and Shinto Eguchi$^{3}$\\[2ex]
\small $^1$Institute of Epidemiology and Preventive Medicine,
National
Taiwan University, Taiwan\\
\small $^2$Institute of Statistical Science, Academia Sinica,
Taiwan\\
\small $^3$Institute of Statistical Mathematics, Japan}
\date{}
\maketitle

\begin{abstract}
Principal component analysis (PCA) is one of the most popular dimension reduction methods. The usual PCA is known to be sensitive to the presence of outliers, and thus many robust PCA methods have been developed.
Among them, the Tyler's M-estimator is shown to be the most robust scatter estimator under the elliptical distribution. However, when the underlying distribution is contaminated and deviates from ellipticity, Tyler's M-estimator might not work well.
In this article, we apply the semiparametric theory to propose a robust semiparametric PCA. The merits of our proposal are twofold. First, it is robust to heavy-tailed elliptical distributions as well as robust to non-elliptical outliers. Second, it pairs well with a data-driven tuning procedure, which is based on active ratio and can adapt to different degrees of data outlyingness.
Theoretical properties are derived, including the influence functions for various statistical functionals and asymptotic normality. Simulation studies and a data analysis demonstrate the superiority of our method.

\end{abstract}

\newpage

\section{Introduction}\label{sec.introdcution}

We consider the problem of robust principal component analysis (PCA) under a contaminated elliptical distribution. A random vector $X\in \mathbb{R}^p$ is said to be elliptically distributed if its pdf takes the form of
\begin{eqnarray}
f(x)=|V_0|^{-1/2}\psi\left(d(x,\mu_0, V_0)\right),\quad x\in \mathbb{R}^p,\label{elliptical}
\end{eqnarray}
where $\mu_0$ is the mean vector, $V_0$ is a positive definite scatter matrix,  $\psi(\cdot)>0$ is an arbitrary radial function, and $d(x,\mu_0, V_0)=(x-\mu_0)^\top V_0^{-1}(x-\mu_0)$. Let the eigenvalue decomposition (EVD) of $V_0$ be denoted by
\begin{eqnarray}
V_0=\sum_{j=1}^{p}\lambda_j\gamma_j\gamma_j^\top, \label{eigen}
\end{eqnarray}
where eigenvalues are assumed distinct $\lambda_1>\lambda_2>\cdots>\lambda_p$. The leading eigenvectors $\gamma_j$'s are the main interest for PCA subspace. The ordinary PCA is conducted based on the EVD of the sample covariance matrix. It is known that the sample covariance matrix is sensitive to the presence of outliers, which can further make the subsequent PCA unreliable.

To guard against outliers, a common strategy is to implement PCA based on a robust scatter estimator. There exist many robust scatter estimators, including M-estimator (Maronna, 1976), S-estimator (Davies, 1987), $\tau$-estimator (Lopuha\"{a}, 1991), Rocke's estimator (Rocke, 1996), and MM-estimator (Tatsuoka and Tyler, 2000;  Salibi{\'a}n-Barrera, Van Aelst and Willems, 2006) among many others. We refer the reader to Maronna and Yohai (2017) for a thorough review and comparison of existing methods. All the above-mentioned robust scatter estimators correspond to a solution pair $(\mu,V)$ satisfying the following system of estimating equations:
\begin{eqnarray}
 \mu &=& \frac{\int w\left(d(x,\mu,V)\right) xdF}{\int w\left(d(x,\mu,V)\right)dF}, \nonumber\\
V&=&c_\psi\int w\left(d(x,\mu,V)\right) (x-\mu)^{\otimes 2} dF,\label{estimating_equation_G}
\end{eqnarray}
where $c_\psi$ is a $\psi$-dependent scalar, $w(\cdot)\ge 0$ is a certain weight function, $a^{\otimes 2}$ stands for $aa^\top$, $F$ is the cdf of $f$, and $dF$ stands for $dF(x)$ for notational simplicity. The sample version of $(\mu,V)$ is obtained by replacing $F$ with the empirical cdf $\hat F$ of the sample $\{X_i\}_{i=1}^n$. The weight function $w(\cdot)$ is method-dependent which aims to downweigh suspicious sample points to achieve robustness. The constant $c_\psi$ depends on $\psi(\cdot)$ only and plays an important role for the consistency of $V$ as an estimator for $V_0$. While $\psi(\cdot)$ plays a role in the construction of $V$, it is also known that the covariance matrix of $X$ under elliptical distribution takes the form ${\rm cov}(X)=a_{\psi}V_0$ for some constant $a_{\psi}$ that depends on $\psi(\cdot)$. As a result, the knowledge of $\psi(\cdot)$ is not critical to PCA eigenvectors, since $V_0$ and $a_{\psi}V_0$ have the same ordering of eigenvalues and share the same eigenvectors. In view of this point, a scatter estimator that does not require the knowledge of $\psi(\cdot)$ is preferred if eigenvectors are of the main interest.

The pioneer scatter estimator without the need to specify $\psi(\cdot)$ is the Tyler's M-estimator (TME) (Tyler, 1987). For a given location estimate $\mu$, TME is defined to be the solution $V_{\rm T}$ of
\begin{eqnarray}
V_{\rm T}&=&p\int \frac{(x-\mu)^{\otimes 2}}{d(x,\mu,V_{\rm T})}dF. \label{estimating_equation.Tyler}
\end{eqnarray}
The robustness of TME to the specification of $\psi(\cdot)$ can partly be seen from its {\it scale-invariance}. If $V_{\rm T}$ is a solution to~\eqref{estimating_equation.Tyler}, so is $aV_{\rm T}$ for any $a>0$. As a result, $V_{\rm T}$ is very robust to heavy tails in an elliptical distribution (named {\it elliptical outliers} hereafter). TME is shown to be the ``most robust'' scatter estimator in the family of elliptical distributions (Tyler, 1987). TME is thus widely used for robust PCA, and has inspired many research works to improve its performance (Ollila and Tyler, 2014; Ollila, Palomar, and Pascal, 2020; Goes, Lerman, and Nadler, 2020).
By noting that the multivariate Kendall's tau matrix and $V_0$ share the same eigenvectors (Marden, 1999), Han and Liu (2018) proposed the elliptical component analysis (ECA) for robust estimation of eigenvectors of $V_0$. TME and ECA can be regarded as semiparametric estimators of $V_0$ under model~(\ref{elliptical}), where $(\mu_0,V_0)$ are the parameters of interest, and $\psi(\cdot)$ is the infinite-dimensional nuisance parameter. Also from a semiparametric viewpoint, Hallin, Oja, and Paindaveine (2006) proposed a semiparametric efficient R-estimator for $V_0$. Later, an extension to the case of complex elliptical distribution and a computational efficient algorithm for implementing R-estimator were developed (Fortunati, Renaux, and Pascal, 2020; Fortunati, Renaux, and Pascal, 2022). The idea to construct R-estimator is also adopted in Hallin, Paindaveine, and Verdebout (2014) to propose an efficient R-estimation for the eigenvectors of $V_0$ directly.

While the above mentioned semiparametric methods are robust to any specific form of $\psi(\cdot)$, their performance relies on the ellipticity of $X$. As a result, biased PCA results can be inevitably encountered under the Huber's $\pi$-contamination model
\begin{eqnarray}\label{FH}
(1-\pi)F +\pi H,
\end{eqnarray}
where $F$ is the cdf of (\ref{elliptical}), $H$ is an arbitrary contamination distribution, and $\pi$ is the contamination proportion (named {\it non-elliptical outliers} hereafter). To mitigate the effect of $H$ under~(\ref{FH}), Chen, Gao, and Ren (2018) defined the matrix depth function, and proposed to use the deepest matrix (i.e., the one with the minimum matrix depth) as a robust estimator for $V_0$. The main contribution of their work is the derivation of the minimax rates for
robust covariance matrix estimation under~(\ref{FH}), which
can be achieved by optimizing over the proposed matrix depth function. However, they also mentioned in their paper that their study is mainly of theoretical interest and the proposed estimators based on matrix depth are challenging to compute.

The aim of this paper is to propose a robust PCA method that can well address the above mentioned robustness issues (including robust to any specific $\psi$ and under Huber's $\pi$-contamination) and computational issues (including feasible implementation and data-driven tuning procedure) in a balanced way. Our proposal is based on model~(\ref{FH}), where $(\mu_0,V_0)$ is the parameter of interest and $(\psi,\pi,H)$ is treated as nuisance parameter, which we call {\it semiparametric PCA} (SPPCA). SPPCA has the following attractive features:

\begin{itemize}
\item[(a)]
The robust scatter estimator of SPPCA is a re-weighted version of TME, and possesses the same scale-invariance property with TME. SPPCA is thus robust to the specification of $\psi(\cdot)$, and hence, to heavy-tailed elliptical outliers.

%We apply the semiparametric theory to construct a class of estimating equations for %$V_0$ by treating $\psi(\cdot)$ as the nuisance parameter. The resulting class of %estimating equations is thus robust to elliptical outliers, i.e., heavy-tailed %elliptical distributions.

\item[(b)]
The hard-threshold weight (\ref{hardthreshold_weight}) ensures the consistency of SPPCA even under the Huber's $\pi$-contamination model~(\ref{FH}). SPPCA is thus robust to non-elliptical outliers.

%makes SPPCA a trimming-based PCA, where a portion of samples are assigned zero weight %to achieve robustness.

%Second, we show that a subclass of the estimating equations can lead to bounded and %vanishing influence functions for eigenvalues and eigenvectors, and thus can be very %robust to the presence of non-elliptical outliers. Our SPPCA will then adopt a %specific choice from this subclass using the hard-threshold weight function %(\ref{hardthreshold_weight}).

\item[(c)]
The hard-threshold weight (\ref{hardthreshold_weight}) connects the tuning parameter of SPPCA with active ratio (AR), which is the proportion of samples being assigned with positive weights. This AR viewpoint enables us to develop a diagnostic tool to help tuning SPPCA without requiring prior knowledge of $F$.

\item[(d)]
SPPCA is implemented via the fixed-point algorithm with little computing effort. SPPCA is thus scalable to the case of large $p$.

\end{itemize}

%{\red The resulting robust scatter estimator is shown to be a re-weighted version of %TME. It is this weight function that makes SPPCA more superior than TME to guard against %non-elliptical outliers. Moreover, the hard-threshold weight makes SPPCA a %trimming-based robust PCA method (i.e., a portion of suspicious samples are assigned %zero weights to achieve robustness). As will be shown later, the hard-threshold weight %(\ref{hardthreshold_weight}) is the core to ensure the consistency of SPPCA even under
%the contaminated distribution (\ref{FH}).}

%Most robust PCA methods involve a certain tuning parameter controlling the tradeoff %between efficiency and robustness. A common approach for tuning parameter selection is %to control the efficiency loss to be under a pre-specified level. To calculate the %efficiency, it relies on the knowledge of $F$, or at least relies on some parametric %assumption on $F$, and is not applicable in the current semiparametric setting. To %resolve this issue, we develop an active ratio (AR) plot to help determining the tuning %parameter of SPPCA. This AR plot is able to reflect the outlyingness of the underlying %data without parametric assumption on $F$, and can be efficiently calculated even for %the case of large $p$. These nice features facilitate the implementation of SPPCA in %practical applications.

The rest of the article is organized as follows. In Section~2 the SPPCA methodology is proposed and theoretical properties are derived. In Section~3 we propose a tuning procedure for hyperparameter selection. Simulation studies and a data analysis are presented in Sections~4-5 to demonstrate the performance of SPPCA. The article ends with a discussion in Section~6. All proofs are placed in Appendix.

%Our major contributions are summarized below.
%\begin{itemize}
%\item
%Derive the semiparametric theory for PCA. Theorem~\ref{thm.nuisance} gives the %orthogonal complement of the nuisance tangent space, which characterizes the %construction of consistent estimation.
%\item
%Derive the influence functions (IFs) for eigenvalues and eigenvectors %(Theorems~\ref{thm.IF.V} and~\ref{thm.V.eigen_IF}) and associated asymptotic normality %(Theorem~\ref{thm.asymptotic_distribution.V}).
%\item
%Propose robust semiparametric estimators for eigenvalues and eigenvectors based on the %robust covariance estimate~(\ref{estimating_equation.V_hat}) and with weight %function~(\ref{hardthreshold_weight}). The proposed estimators enjoy the good property %of having bounded IFs. Implementation details, including tuning parameter selection, are %provided (Sections~\ref{sec.fixed_point} and~\ref{sec.tuning}).
%\end{itemize}}

\section{Methodology}

\subsection{Semiparametric estimation}\label{sec.semi_PCA}

Model (\ref{elliptical}) suggests a semiparametric framework under the family of elliptical distributions, where $(\mu_0, V_0)$ are the parameters of interest, and $\psi(\cdot)$ is the infinite-dimensional nuisance parameter. Let $\mathcal{H}=\{g(X)\in L_2: E\{g(X)\}=0\}$ be the $L_2$ Hilbert space of mean zero random functions of $X$. The {\it nuisance tangent space}, denoted by $\Lambda_{\psi}$, is defined to be the closure of all nuisance tangent spaces of parametric models for $\psi(\cdot)$. The semiparametric theory then states that any regular asymptotic linear estimator of $(\mu_0,V_0)$ associates with a unique influence function in $\mathcal{H}$, which belongs to the orthogonal complement of $\Lambda_{\psi}$ (Tsiatis, 2007). Thus, this orthogonal complement $\Lambda_{\psi}^{\perp}$ characterizes the construction of consistent estimators for $V_0$. We have the following result regarding the semiparametric estimation of $(\mu_0,V_0)$.

\begin{theorem}[semiparametric estimation for $(\mu_0,V_0)$]\label{thm.nuisance}
Assume the semiparametric model~(\ref{elliptical}), where $(\mu_0,V_0)$ are the parameters of interest and $\psi(\cdot)$ is the infinite-dimensional nuisance parameter. Then,  we have
\begin{eqnarray*}
\Lambda_{\psi}^\perp = \Big\{ g(X)\in L_2: E\{g(X)|d(X,\mu_0,V_0)\}=0 \Big\}.
\end{eqnarray*}
Moreover, for any $w(\cdot)\ge 0$ such that $E\left\|w\left(d(X,\mu_0,V_0)\right)(X-\mu_0)^{\otimes 2} \right\|_{\rm F}<\infty$, we have
\begin{eqnarray*}
&&w\left(d(X,\mu_0,V_0)\right) \left[\begin{array}{ll}
\quad\quad \quad\quad\quad\quad X-\mu_0\\
\vec\Big\{(X-\mu_0)^{\otimes 2}-\frac{d(X,\mu_0,V_0)}{p}V_0\Big\}
\end{array}\right]\in\Lambda_{\psi}^\perp.
\end{eqnarray*}
\end{theorem}

In the rest of discussion, $w(\cdot)$ is assumed to be a non-increasing function satisfying the condition stated in Theorem~\ref{thm.nuisance}. Theorem~\ref{thm.nuisance} ensures that $(\mu_0,V_0)$ is a solution pair to the system of estimating equations:
\begin{eqnarray}\label{estimating_equation.V_original}
&&\int w\left(d(x,\mu,V)\right)(x-\mu)dF=0\nonumber\\
&&\int w\left(d(x,\mu,V)\right)\left\{(x-\mu)^{\otimes 2}-\frac{d(x,\mu,V)}{p}V\right\}dF=0,
\end{eqnarray}
which can be equivalently expressed as
\begin{eqnarray}
 \mu &=& \frac{\int w\left(d(x,\mu,V)\right) xdF}{\int w\left(d(x,\mu,V)\right)dF}, \label{estimating_equation.mu} \\
 V &=&  \frac{p\int w\left(d(x,\mu,V)\right)(x-u)^{\otimes 2} d F}
{\int w\left(d(x,\mu,V)\right)d(x,\mu,V) d F}. \label{estimating_equation.V}
\end{eqnarray}
The following theorem ensures that, at the population level, (\ref{estimating_equation.V}) also possesses the same scale-invariance property as TME has.

\begin{theorem}[solution set]\label{thm.shape_uniqueness}
Assume the elliptical distribution~(\ref{elliptical}). Let $\mathcal{V}_0=\{aV_0:a>0\}$ be the set of matrices that are the same as $V_0$ up to a positive multiplicative scalar.
\begin{itemize}
\item[(a)]
$\mu=\mu_0$ and any $V \in \mathcal{V}_0$ constitute a solution pair for (\ref{estimating_equation.mu})-(\ref{estimating_equation.V}).

\item[(b)]
If $\psi(\cdot)$ is monotone, then the solutions of (\ref{estimating_equation.mu})-(\ref{estimating_equation.V}) must take the form $\mu=\mu_0$ and $V\in\mathcal{V}_0$.
\end{itemize}
\end{theorem}

\noindent Although the solutions of the semiparametric estimating equations~(\ref{estimating_equation.mu})-(\ref{estimating_equation.V}) are not unique, Theorem~\ref{thm.shape_uniqueness} ensures that any element in $\mathcal{V}_0$ (i.e., having the same shape with $V_0$) must be a solution. This is sufficient for the validity of conducting PCA based on $\mathcal{V}_0$, since $V_0$ and $aV_0$ have the same eigenvectors. Our proposed robust PCA procedure is then based on the solution set given by~(\ref{estimating_equation.V}), which we call {\it semiparametric PCA} (SPPCA).

The robustness of SPPCA is twofold. First, (\ref{estimating_equation.mu})-(\ref{estimating_equation.V}), evaluated at $(\mu,V)=(\mu_0,aV_0)$, are valid for arbitrary $\psi(\cdot)$. That is, it is robust to any specific choice of $\psi(\cdot)$ in the family of elliptical distributions. This robustness can be expected, since our procedure is based on semiparametric theory with $\psi(\cdot)$ being treated as the nuisance parameter. Thus, SPPCA is robust to the presence of elliptical outliers (i.e., heavy tails). Furthermore, as will be shown later, (\ref{estimating_equation.mu})-(\ref{estimating_equation.V}) can be robust to non-elliptical outliers (i.e., when $X$ is generated from (\ref{FH})), provided that $w(\cdot)$ is properly chosen. We assume that $w(\cdot)$ is given for the moment, and will discuss its choice later in Section~\ref{sec.w}.

Note that equation (\ref{estimating_equation.V}) can be equivalently expressed as
\begin{eqnarray}\label{est_eq.weighted_Tyler}
V=p\int  \left\{\frac{h(d(x,\mu,V))}{\int h(d(x,\mu,V))d F}\right\}\frac{(x-\mu)^{\otimes 2}}{d(x,\mu,V)}dF\quad {\rm with }\quad h(u) = w(u)u.
\end{eqnarray}
It can be seen that (\ref{est_eq.weighted_Tyler}) is a re-weighted version of TME with the weight $h(d(x,\mu,V))$, and $V_{\rm T}$ in (\ref{estimating_equation.Tyler}) corresponds to the choice of $h(u)=1$.
While both $V$ and $V_{\rm T}$ are robust to the form of $\psi(\cdot)$, the performance of $V_{\rm T}$ is not guaranteed under the Huber's $\pi$ contamination model~(\ref{FH}). In this situation, $V_{\rm T}$ still assigns equal weight to each sample point generated from either $F$ or $H$, and the eigenvectors of $V_{\rm T}$ can be very different from the target eigenvectors of $V_0$. This failure of $V_{\rm T}$ under the contamination by $H$ is mainly due to its ignorance of the length information of $(X-\mu)$. Note that $V_{\rm T}$ depends on $(X-\mu)$ only through $\{d(X,\mu,V_{\rm T})\}^{-1/2}(X-\mu)$. The length of $(X-\mu)$, however, contains useful information to help separating $F$ from $H$. This information can be properly utilized in $V$ via $h(\cdot)$ as shown in (\ref{est_eq.weighted_Tyler}).

\begin{remark}\label{rmk.tyler}
From the re-weighted expression (\ref{est_eq.weighted_Tyler}), the weight $\frac{h(d(X,\mu,V))}{\int h(d(x,\mu,V))dF}$ depends on $X$ over the entire distribution $F$. Hence, SPPCA cannot be expressed as a special case of (\ref{estimating_equation_G}), unless $h(u)=1$. On the other hand, TME is a special case of (\ref{estimating_equation_G}) with $c_\psi=1$ and $w(u)=p/u$. This view point indictes the difference between SPPCA and TME.
\end{remark}

Recall the empirical distribution function $\widehat F$ of the sample, and define
\begin{eqnarray}\label{estimating_equation.V_hat}
\widehat{\mathcal{V}}=\left\{\widehat V:~\widehat\mu=\frac{\int w(d(x,\widehat\mu,\widehat V))xd\widehat F}{\int w(d(x,\widehat\mu,\widehat V))d\widehat F},\quad\widehat V=
\frac{p\int w(d(x,\widehat\mu,\widehat V))(x-\widehat\mu)^{\otimes 2} d\widehat F}{\int w(d(x,\widehat\mu,\widehat V))d(x,\widehat\mu,\widehat V) d\widehat F}\right\}.
\end{eqnarray}
The set $\widehat{\mathcal{V}}$ consists of all $\widehat V$ such that $(\widehat\mu,\widehat V)$ is a solution of the sample version of  (\ref{estimating_equation.mu})-(\ref{estimating_equation.V}). The validity of $\widehat{\mathcal{V}}$ is ensured by Theorem~\ref{thm.shape_uniqueness}, which implies that any $\widehat V\in \widehat{\mathcal{V}}$ produces consistent eigenvectors estimation under model~(\ref{elliptical}). Different from $\mathcal{V}_0$, however, the elements of $\widehat{\mathcal{V}}$ are not necessarily proportional to each other (i.e., they do not have the same shape of scatter), since $\widehat F$ is not elliptically symmetric in empirical data. This introduces an issue for SPPCA to selecting an appropriate element from $\widehat{\mathcal{V}}$ to perform PCA. This issue will be discussed in Section~\ref{sec.tuning}.

\subsection{Influence functions}\label{sec.robustness}

Directly investigating the statistical properties of SPPCA via (\ref{estimating_equation.mu})-(\ref{estimating_equation.V}) is not straightforward, since only the shape of $V_0$ is identifiable in the solution set $\mathcal{V}_0$, but not the scale. Thus, we will study the influence functions (IFs) of eigenvectors and ratios of eigenvalues that are invariant with respect to the scale change.

Firstly, solutions given by (\ref{estimating_equation.mu})-(\ref{estimating_equation.V}) can be made unique by fixing the scale size to unity. Let $s(\cdot)$ be a given positive {\it scale function} satisfying $s(aV_0)=as(V_0)$ for any $a>0$, e.g., $s(V_0)=|V_0|^{1/p}$. It gives a unique shape matrix $V_{s0}\in\mathcal{V}_0$ having unit scale $s(V_{s0})=1$, where $V_{s0}$ is given by
\begin{eqnarray}\label{Sigma0_reparameterization}
V_{s0}=\frac{1}{\sigma_{s0}}V_0\quad {\rm with}\quad\sigma_{s0}=s(V_0).
\end{eqnarray}
The corresponding eigenvalue decomposition is $V_{s0}=\sum_{j=1}^{p}\lambda_{sj}\gamma_j\gamma_j^\top$ with $\lambda_{sj}=\lambda_j/\sigma_{s0}$.
Let $\mu_s(F)$ and $V_s(F)$ be the functionals of the solution to the constrained version of (\ref{estimating_equation.mu})-(\ref{estimating_equation.V}) with constraint $s(V)=1$. That is, $V_s=V/{s(V)}$, and thus $s(V_s)=1$.
We will investigate the statistical properties of SPPCA via the scale-constrained version $V_s(F)$ as a function of the underlying distribution $F$.

The robustness of SPPCA can be quantitatively evaluated via the IF. The IF of a statistical functional $T(F)$ is defined to be ${\rm IF}_T(x;F)=\frac{\partial}{\partial\varepsilon}T((1-\varepsilon)F+\varepsilon\delta_x)|_{\varepsilon=0}$, where $\delta_x$ is the point mass at $x$. It measures the instantaneous change of $T(\cdot)$ when the distribution $F$ is perturbed by $\delta_x$. We have the following results.

\begin{theorem}[IFs for the location vector and scatter matrix] \label{thm.IF.V}
The IFs of $\mu_s(F)$ and $V_s(F)$ are determined by
\begin{eqnarray*}
&&\IF_{\mu_s}(x;F) = \eta_s w(d(x,\mu_0,V_{s0}))(x-\mu_0)\\
&&\IF_{V_s}(x;F)-\frac{\tr\{\IF_{V_s}(x;F)V_{s0}^{-1}\}}{p}V_{s0} = \phi_s w(d(x,\mu_0,V_{s0})) \left\{(x-\mu_0)^{\otimes 2}-\frac{d(x,\mu_0,V_{s0})}{p}V_{s0}\right\},
\end{eqnarray*}
where $\eta_s=\{\frac{2}{p}\int (y^\top y)w(y^\top y)\psi_s'(y^\top y)dy\}^{-1}$ and $\phi_s=\{\frac{2}{p(p+2)}\int (y^\top y)^2w(y^\top y)\psi_s'(y^\top y)dy\}^{-1}$ with $\psi_s(u)=\sigma_{s0}^{-p/2}\psi(u/\sigma_{s0})$.
\end{theorem}

\noindent The exact form of $\IF_{V_s}(x;F)$ depends on the choice of $s(\cdot)$. The forms of the IFs of eigenvalue ratios and eigenvectors, however, can be determined regardless the choice of $s(\cdot)$. Recall that $V_{s0}=\sum_{j=1}^{p}\lambda_{sj}\gamma_j\gamma_j^\top$.

\begin{theorem}[IFs for eigenvalue ratios and eigenvectors]\label{thm.V.eigen_IF}
Let $\lambda_{sj}(F)$ and $\gamma_{sj}(F)$ be the $j^\th$ eigenvalue and associated eigenvector of the scaled scatter matrix $V_s(F)$ and let $\lambda_{sij}(F)=\lambda_{sj}(F)/\lambda_{si}(F)$ be the ratio of the $i^\th$ eigenvalue to the $j^\th$ eigenvalue.
Then, the IFs of $\gamma_{sj}(F)$ and $\lambda_{sij}(F)$, $1\le i\ne j\le p,$ are given by
\begin{eqnarray*}
\IF_{\lambda_{sij}}(x;F)  &=& \phi_s w\left(d(x,\mu_0,V_{s0})\right) \lambda_{ij}\left[\frac{\{\gamma_j^\top (x-\mu_0)\}^2}{\lambda_{sj}}-\frac{\{\gamma_i^\top (x-\mu_0)\}^2}{\lambda_{si}}\right] ,\\
\IF_{\gamma_{sj}}(x;F)  &=& \phi_s w\left(d(x,\mu_0,V_{s0})\right) \{\gamma_j^\top (x-\mu_0)\}(\lambda_{sj}I_p-V_{s0})^+(x-\mu_0),
\end{eqnarray*}
where $\phi_s$ is defined in Theorem~\ref{thm.IF.V}. Moreover, we have $\|\IF_{\lambda_{sij}}(x;F)\|=O\big\{h(d(x,\mu_0,V_{s0}))\big\}$ and
$\|\IF_{\gamma_{sj}}(x;F)\|=O\big\{h(d(x,\mu_0,V_{s0}))\big\}$, where $h(\cdot)$ is defined in (\ref{est_eq.weighted_Tyler}).
\end{theorem}

\noindent One implication of Theorem~\ref{thm.V.eigen_IF} is that the robustness of SPPCA is controlled by $h(\cdot)$, which plays an important role in bounding the norm of IFs. This observation helps us to select an appropriate weight function $w(\cdot)$, which will be discussed in the next subsection. With Theorem~\ref{thm.V.eigen_IF}, the asymptotic properties of SPPCA can be established as stated  below.

\begin{theorem}[asymptotic normality]\label{thm.asymptotic_distribution.V}
Let  $\widehat\lambda_{sj}$ and $\widehat\gamma_{sj}$ be the $j^{\rm th}$ eigenvalue and associated eigenvector of $\widehat V_s$, where $\widehat V_s =V_s(\widehat F)$. Let $\widehat\lambda_{sij}=\widehat\lambda_{sj}/\widehat\lambda_{si}$, and $Q_0=I-\frac{1}{p}\vec(V_0)\vec(V_0^{-1})^\top$. Then, we have the weak convergence $Q_0\sqrt{n}\vec(\widehat V_s-V_{s0})\stackrel{d}{\to}N(0,\Sigma_{V_{s0}})$, $\sqrt{n}(\widehat \gamma_{sj}-\gamma_j)\stackrel{d}{\to}N(0,\Sigma_{\gamma_j})$, and $\sqrt{n}(\widehat \lambda_{sij}-\lambda_{ij})\stackrel{d}{\to}N(0,\Sigma_{\lambda_{ij}})$  as $n\to\infty$, where
\begin{eqnarray*}
\Sigma_{V_{s0}} &=& \xi_s\left\{(I+K_p)(V_{s0}\otimes V_{s0})-\frac{2}{p}\vec(V_{s0})\vec(V_{s0})^\top\right\}\\
\Sigma_{\gamma_j} &=& \xi_s\lambda_{j}V_{0}(\lambda_{j}I-V_{0})^{+2} \\
\Sigma_{\lambda_{ij}} &=& 4\xi_s\lambda_{ij}^2
\end{eqnarray*}
with $\xi_s = \frac{\phi_s^2}{p(p+2)}\int (y^\top y)^2w^2(y^\top y)\psi_s(y^\top y)dy$, $\psi_s(u)=\sigma_{s0}^{-p/2}\psi(u/\sigma_{s0})$ and  $(\cdot)^{+2}$ denoting the square of the Moore-Penrose pseudoinverse.
\end{theorem}

We have investigated the statistical properties of SPPCA via the scaled version $V_s(\cdot)$ without specifying an explicit form for $s(\cdot)$. The choice of $s(\cdot)$ seems to be an issue when implementing SPPCA. It can be seen form Theorem~\ref{thm.asymptotic_distribution.V} that $\Sigma_{\gamma_j}$ and $\Sigma_{\lambda_{ij}}$ depend on $s(\cdot)$ only through $\xi_s$. Thus, one might consider the optimal $s(\cdot)$ such that $\xi_s$ is minimized. The optimal scale function, however, depends on the unknown $\psi(\cdot)$ and can be very difficult to obtain it. Fortunately, the form of $s(\cdot)$ can be left unspecified when implementing SPPCA. This can be seen from the fact that $\widehat V_s\in \widehat{\mathcal{V}}$ for any choice of  $s(\cdot)$. Thus, one convenient idea to implement SPPCA is to construct $\widehat{\mathcal{V}}$ directly, and then select an element of $\widehat{\mathcal{V}}$ to conduct PCA (i.e., selecting an element from $\widehat{\mathcal{V}}$ implicitly determines a choice of $s(\cdot)$). Detailed implementation algorithm and an informative selection procedure for $\widehat{\mathcal{V}}$ will be discussed in the subsequent discussion.

\subsection{The choice of $w(\cdot)$}\label{sec.w}

From Theorem~\ref{thm.V.eigen_IF}, we see that $h(\cdot)$ plays an important role in controlling the order of magnitude for influence functions. It suggests that a robust SPPCA procedure should associate with a weight function $w(\cdot)$ such that
\begin{equation}\label{h_conditions}
\sup_uh(u)<\infty~~{\rm and}~~ \lim_{u\to\infty}h(u)=0.
\end{equation}
These two conditions ensure that $\|\IF_{\lambda_{sij}}(x;F)\|$ and $\|\IF_{\gamma_{sj}}(x;F)\|$ are bounded and that $\lim_{\|x\|\to\infty}\|\IF_{\lambda_{sij}}(x;F)\|=0$ and $\lim_{\|x\|\to\infty}\|\IF_{\gamma_{sj}}(x;F)\|=0$.
To achieve (\ref{h_conditions}), we consider the following hard-threshold exponential weight
\begin{equation}
w(u) = e^{-u}\cdot\mathcal{I}\left\{ e^{-u}>\alpha\right\},
\label{hardthreshold_weight}
\end{equation}
where $\mathcal{I}\{\cdot\}$ is the indicator function, and $\alpha\in(0,1)$ is a pre-specified hard-threshold value. We set $\alpha=0.05$ throughout this article.
As will be seen later that the scale of $u$ plays an important role for the performance of weighted scatter estimation, while the choice of $\alpha$ is not so crucial. It is also possible to use other weight functions satisfying conditions~(\ref{h_conditions}). Note that the indicator function $\mathcal{I}\left\{ e^{-u}>\alpha\right\}$ in (\ref{hardthreshold_weight}) introduces a trimming mechanism (i.e., a portion of samples will be assigned with zero weights) for SPPCA. The idea of using a trimming mechanism to achieve robustness can also be found in the literature, e.g., Rocke (1996) and Croux et al. (2017), among others.
The benefit of using (\ref{hardthreshold_weight}) to implement SPPCA is rigorously shown below. The proof is a direct consequence of Theorem~\ref{thm.shape_uniqueness} and is omitted.

\begin{corollary}\label{thm.estimating_equation.V.hard_threshold}
Under the weight function (\ref{hardthreshold_weight}), the estimating equation (\ref{estimating_equation.V_original}) becomes
\begin{eqnarray}
&&\int_{\mathcal{B}(\mu,V)}e^{-d(x,\mu,V)}(x-\mu)dF=0,\nonumber\\
&&\int_{\mathcal{B}(\mu,V)}e^{-d(x,\mu,V)}\left\{(x-\mu)^{\otimes 2}-\frac{d(x,\mu,V)}{p}V\right\}dF=0,\label{estimating_equation.V.w}
\end{eqnarray}
where $\mathcal{B}(\mu,V)$ is a subset of support of $F$ and is given by
\begin{eqnarray*}
\mathcal{B}(\mu,V)=\left\{x: e^{-d(x,\mu,V)}>\alpha\right\}=\left\{x:d(x,\mu,V)<\ln(\alpha^{-1})\right\}.
\end{eqnarray*}
Moreover, $\mu=\mu_0$ and $V\in\mathcal{V}_0$ are the solutions of (\ref{estimating_equation.V.w}).
\end{corollary}

From Corollary~\ref{thm.estimating_equation.V.hard_threshold}, we have that $(\mu_0,aV_0)$ is a solution of (\ref{estimating_equation.V.w}) for any given $a>0$.
It also implies that trimming sample points outside $\mathcal{B}(\mu_0,aV_0)$ is not harmful to the consistency of SPPCA. Note that $V$ with a smaller scale size results in a smaller volume of $\mathcal{B}(\mu,V)$. These observations show that a consistent PCA procedure for eigenvectors is achievable in the presence of contamination distribution $H$, provided that $a$ is properly chosen such that $\mathcal{B}(\mu_0,aV_0)$ contains no outliers. This requirement is formally stated in the following definition. Let $P_H(\cdot)$ denote the probability function under $H$.

\begin{definition}[centrally separable]\label{def.separable}
An elliptical distribution $F$, having pdf given by~(\ref{elliptical}), is said to be centrally $a^*$-separable from $H$, if $P_{H}\{X\in \mathcal{B}(\mu_0,aV_0)\}=0$ for some $a>0$, and $a^*$ is the maximum value of such $a$, i.e., $a^*=\sup\big\{a:P_{H}\{X\in \mathcal{B}(\mu_0,aV_0)\}=0\big\}$.
\end{definition}

\noindent The central separability condition does not necessarily require the support of $F$ to be entirely separable from the support of $H$. It requires the existence of a region around the center $\mu_0$ of $F$ such that this region is not contaminated by $H$, i.e., $P_{H}\{X\in \mathcal{B}(\mu_0,aV_0)\}=0$ for some $a>0$ (see also Remark~\ref{rmk.separable} for further discussion). In this situation, we have a chance to recover the shape of $V_0$ via solving (\ref{estimating_equation.V.w}), even under the Huber's $\pi$-contamination model~(\ref{FH}). In the rest of this article, SPPCA is equipped with the hard-threshold exponential weight~(\ref{hardthreshold_weight}) unless it is otherwise stated.

\begin{theorem}[robustness to contamination]\label{thm.consistency.contaminated}
Assume $F$ is centrally $a^*$-separable from $H$.
Under the Huber's $\pi$-contamination model~(\ref{FH}), we have that $\mu=\mu_0$ and $V\in \{aV_0:0<a\le a^*\}$ are the solutions of (\ref{estimating_equation.V.w}).
\end{theorem}

\noindent
Theorem~\ref{thm.consistency.contaminated} implies that, estimators obtained from (\ref{estimating_equation.V.w}) still have the desirable property of Fisher consistency under~(\ref{FH}), provided that the scale size $a$ is small enough ($\le a^*$). This also indicates that elements in $\widehat{\mathcal{V}}$ with a smaller scale size has a better potential to produce a consistent PCA procedure in the presence of outliers. Detailed selection procedure for $\widehat{\mathcal{V}}$ will be discussed in Section~\ref{sec.tuning}.

We remind the reader that the results of Theorems~\ref{thm.IF.V}-\ref{thm.asymptotic_distribution.V} are valid for any $w(\cdot)$, and are not limited to the hard-threshold exponential weight (\ref{hardthreshold_weight}). The choice of (\ref{hardthreshold_weight}), however, supports the success of SPPCA to recover the shape of $V_0$ under the Huber's $\pi$-contamination model~(\ref{FH}) as shown in Theorem~\ref{thm.consistency.contaminated}.

\begin{remark}\label{rmk.separable}
The central $a^*$-separability condition implicitly requires the support of $H$ to be bounded, which seems to be a strong assumption. It is possible to relax this assumption by requiring
$P_{H}\{X\in \mathcal{B}(\mu_0,aV_0)\}$ converges to 0 at a certain rate such that its influence on $F$ is ignorable. For the sake of simplicity,
we still use Definition~\ref{def.separable} as our working assumption to facilitate the development of SPPCA.
\end{remark}

\subsection{Implementation}\label{sec.fixed_point}

In this subsection, we provide an implementation algorithm for obtaining a solution set $\widehat{\cal V}$. With $\widehat{\cal V}$, a specific choice of $\widehat V$ is picked from the set (see Algorithm~2 in Section~\ref{sec.tuning}) for final eigenvector estimation.
The implementation of SPPCA has the same simplicity as the fixed-point algorithm for TME. Starting from an initial value $(\mu_{\rm ini}, V_{\rm ini})$ and using equation~(\ref{estimating_equation.V_hat}), the iterative update is given by
\begin{eqnarray}\label{fixed_point_algorithm}
\mu_{(k+1)}= \frac{\int w(d(x,\mu_{(k)},V_{(k)}))xd\widehat F}{\int w(d(x,\mu_{(k)},V_{(k)}))d\widehat F},\quad V_{(k+1)}= \frac{p\int w(d(x,\mu_{(k)},V_{(k)}))(x-\mu_{(k)})^{\otimes 2} d\widehat F}
{\int w(d(x,\mu_{(k)},V_{(k)}))d(x,\mu_{(k)},V_{(k)}) d\widehat F},
\end{eqnarray}
where our choice of weight function is given in~(\ref{hardthreshold_weight}).

Recall that (\ref{estimating_equation.V}) has solutions of the form $\{aV_0:a>0\}$. This motivates us to construct (a discretized set of) $\widehat {\mathcal{V}}$ by choosing $(\mu_{\rm ini},V_{\rm ini})\in\{(\widetilde{\mu},a\widetilde V):a\in \mathcal{A}\}$, where $\mathcal{A}=\{a_1,a_2,\ldots,a_m\}$ is a pre-determined set of $m$ discretized values for the scale variable~$a$, and $(\widetilde{\mu},\widetilde V)$ is a robust initial estimate of $(\mu_0,V_0)$. We set
\begin{itemize}
\item
$\widetilde \mu$ as the marginal sample median  (i.e., component-wise sample median) of $\{X_i\}_{i=1}^n$,
\item
$\widetilde V$ as a diagonal matrix with the diagonal elements being the component-wise truncated standard deviation of $\{X_{i}\}_{i=1}^n$ using the $\tau$-scale of Yohai and Zamar (1988).
\end{itemize}
The implementation procedure is summarized below.\\

\hrule\vspace{0.05cm} \hrule  \vspace{0.25cm} {\noindent {\bf Algorithm 1.} Implementation for obtaining a solution set $\widehat{\mathcal{V}}$ of scatter matrix} \vspace{0.25cm} \hrule
\begin{itemize}
\item[1.]
Inputs: empirical distribution $\widehat F$, weight function $w(\cdot)$, a robust initial estimator $(\widetilde \mu, \widetilde V)$ described above, and the set $\mathcal{A}$ consisting of candidate values of~$a$.

\item[2.]
Set multiple initials $(\mu_{\rm ini},V_{\rm ini})=(\widetilde{\mu},a\widetilde V)$ for all $a\in \mathcal{A}$. For each initial, do the fixed-point iteration~(\ref{fixed_point_algorithm}), and obtain $\widehat V(a)$ at convergence.

\item[3.]
Output solution set $\widehat{\mathcal{V}}=\left\{\widehat V(a):a\in \mathcal{A}\right\}$.
\end{itemize}
\vspace{-0.1cm}\hrule\vspace{0.05cm} \hrule

\vspace{0.3cm}

The construction of $\widehat{\mathcal{V}}$ relies on using multiple initial values  $V_{\rm ini}\in\{a\widetilde V:a\in \mathcal{A}\}$ to implement the fixed-point algorithm (\ref{fixed_point_algorithm}). Since (\ref{fixed_point_algorithm}) depends on $a$ via $d(x,\mu_{\rm ini},a\widetilde V)=d(x,\mu_{\rm ini},\widetilde V)/a$, a reasonable magnitude of $a$ should be of the order $O(p)$. A larger $a$ will tend to result in a $\widehat V(a)$ having a larger scale size as shown below.

\begin{theorem}[scaling effect of $a$]\label{thm.fixed_point_algorithm}
At the convergence of the fixed-point algorithm~(\ref{fixed_point_algorithm}), we have $|\widehat V(a)|^{1/p}=ac_a^{1/p}|\widetilde V|^{1/p}$ for some constant $c_a\in(0,1)$.
\end{theorem}

\noindent This theorem shows that the size of $a$ controls the scale size of $\widehat V(a)$, which is approximately $a$ times the scale size of the initial scatter matrix $\widetilde V$. From this viewpoint, selecting an element in $\widehat{\mathcal{V}}=\left\{\widehat V(a):a\in \mathcal{A}\right\}$ can be understood as implicitly selecting a suitable scale size $a$. This issue of tuning will be further elucidated in Section~\ref{sec.tuning}.

\begin{remark}\label{rmk.fixed_point_algorithm}
The fixed-point algorithm (\ref{fixed_point_algorithm}) can be unstable when $p\approx n$ or even fail when $p>n$ due to the calculation of $V^{-1}$ in $d(x,\mu,V)$. For the sake of stability and scalability, we suggest to approximate $d(x,\mu,V)$ by $d(x,\mu,D_V)$ when implementing (\ref{fixed_point_algorithm}), where $D_V$ is a diagonal matrix having the same diagonal elements as $V$.
\end{remark}

\section{Informative Self-Tuning of SPPCA}\label{sec.tuning}

The aim of this section is to propose a diagnostic tool to help determining the tuning parameter of SPPCA. Our proposal is based on the active ratio (AR), which is the proportion of active sample points to enter the analysis relative to the whole data sample. We first review some basic ideas of existing AR-based tuning methods in Section~\ref{sec.AR_review}. Next we propose our tuning procedure in Sections~\ref{sec.AR_sppca}-\ref{sec.AR_proposed_method}.

\subsection{Tuning via active ratio}\label{sec.AR_review}

Almost every robust PCA method requires a proper tuning to control the tradeoff between estimation efficiency and robustness. For SPPCA, this corresponds to selecting a scale size $a$ to obtain the shape estimator $\widehat V(a)$ from Algorithm~1. A common criterion is to select a tuning parameter value, where the asymptotic relative efficiency (ARE) reaches a pre-specified level. Usually the multivariate normal distribution is used as the baseline for ARE calculation. However, the target distribution $F$ can deviate from multivariate normal, which makes this ARE-based tuning leading to a biased PCA estimation. Another drawback is that ARE has no direct relationship with $(\pi,H)$, which makes it difficult for practitioners to determine a suitable ARE value.

There is a branch of methods for robust PCA by trimming a portion of suspicious sample points, e.g., the minimum volume ellipsoid (MVE) estimator and the minimum covariance determinant (MCD) estimator (Rousseeuw, 1985), ROBPCA (Hubert, Rousseeuw, and Vanden Branden, 2005), and trimmed-PCA (Croux {\it et al.}, 2017). For these methods, they used AR as the tuning parameter to control the number of data points entering the analysis, and hence, to control the level of robustness. The benefit of the AR-based tuning strategy is that the optimal AR value directly associates with the underlying proportion $\pi$ in the Huber's $\pi$-contamination model~(\ref{FH}). For example, assume that the support of $F$ is well separated from that of $H$. Ideally, samples from $H$ should be discarded so that the analysis is ideally based on samples from $F$ only. Under this ideal case, the optimal AR is $1-\pi$. However, in practice, $\pi$ is unknown and the supports of $F$ and $H$ often are overlapped. In this situation, the optimal AR will be smaller than $1-\pi$, due to a portion of sample points from $H$ cannot be well separated from $F$. This overlapping phenomenon complicates the AR selection and some extra information is required to assist the AR-based tuning strategy.

To guide the AR-based tuning, Paindaveine and Van Bever (2019) developed the {\it Tyler shape depth} of a shape matrix and discussed its application to tuning for MCD estimator. The main idea is as follows. When MCD is properly tuned so that outliers are trimmed out, it should have a relatively large Tyler shape depth for these active sample points. The authors then proposed to visually inspect the plot of Tyler shape depths of MCD (in $y$-axis) versus the AR values (in $x$-axis). When the AR values go beyond a certain threshold, a flat curve, which is formed by relatively large but stable Tyler shape depths, can be expected. That is, a change-point at the threshold can be expected. The Tyler shape depth thus provides some additional information to assist the AR-based tuning.

\subsection{Active ratio of SPPCA}\label{sec.AR_sppca}

The tuning parameter of SPPCA (i.e., $a$ of $\widehat V(a)$) is closely related to AR. By the hard-threshold exponential weight~(\ref{hardthreshold_weight}), the quantities $(\widehat\mu(a), \widehat V(a))$ from Algorithm~1 depend on sample points belonging to $\mathcal{B}(\widehat \mu(a), \widehat V(a))$ only. This motivates us to define the AR value of $\widehat V(a)$ to be
\begin{eqnarray}
{\rm AR}(a)=\frac{1}{n}\sum_{i=1}^{n}\mathcal{I}\left\{X_i\in \mathcal{B}(\widehat\mu(a),\widehat V(a))\right\}.\label{AR}
\end{eqnarray}
This suggests that existing AR-based tuning techniques can be applied for SPPCA. In Theorem~\ref{thm.fixed_point_algorithm}, we see that $a$ can be viewed as a scale factor in $\widehat V(a)$. On one hand, $\widehat V(a)$ with a larger scale size tends to produce a larger ${\rm AR}(a)$. On the other hand, Theorem~\ref{thm.consistency.contaminated} implies that $\widehat V(a)$ with a smaller scale size has a better potential to produce consistent PCA estimation in the presence of outliers. These observations lead us to consider a selection criterion satisfying the following two desired properties:
\begin{itemize}
\item
The value of $a$ should be as large as possible, so that ${\rm AR}(a)$ is large enough to maintain the estimation efficiency of $\widehat V(a)$.

\item
The value of $a$ should satisfy $P_{H}\{X\in \mathcal{B}(\mu_0, aV_0)\}\approx 0$, so that SPPCA based on $\widehat V(a)$ is consistent for eigenvectors estimation under $(1-\pi)F+\pi H$.
\end{itemize}
If $F$ is $a^*$-centrally separable from $H$, then the optimal choice of $a$ is obviously $a=a^*$. In this situation, SPPCA based on $\widehat V(a^*)$ utilizes the most sample points from $F$ without involving sample points from $H$. This also suggests that selecting an element in $\widetilde{\mathcal{V}}$ for SPPCA is equivalent to the estimation of $a^*$. In next subsection, we will discuss a visual inspection method to determine $a^*$ via ${\rm AR}(a)$.

\subsection{An AR-based diagnostic tool to determine $a^*$}\label{sec.AR_proposed_method}

Owing to the connection between $\widehat V(a)$ and ${\rm AR}(a)$ in (\ref{AR}), the  Tyler shape depth (Paindaveine and Van Bever, 2019) can be directly applied to determine $a^*$. This strategy, however, can be problematic when $p$ is large, due to the high computational loading for calculating the Tyler shape depth. To overcome this difficulty, we alternatively propose a scalable (see Remark~\ref{rmk.computation} for details) diagnostic tool to help tuning SPPCA.

Define the AR curve of SPPCA (see Remark~\ref{rmk.AR} for the specification of $\mathcal{A}$) to be
\[\mathcal{C}_{\rm AR}=\{(a,{\rm AR}(a)):a\in \mathcal{A}\}.\]
The curve $\mathcal{C}_{\rm AR}$ contains information about the optimal $a^*$ based on the following observations of a change-point phenomenon. Define the transformation $Z=\frac{-1}{\ln \alpha}d(X,\mu_0,V_0)$ of $X$, and let $\Phi_F(a)=P_F(Z\le a)$ and $\Phi_H(a)=P_H(Z\le a)$. For $F$, which is (approximately) centrally $a^*$-separable from $H$, we have
\begin{eqnarray}
P\{X \in \mathcal{B}(\mu_0, aV_0)\}&=&(1-\pi)P_F\{X \in \mathcal{B}(\mu_0, aV_0)\}+\pi P_H\{X \in \mathcal{B}(\mu_0, aV_0)\}\nonumber\\
&\approx&(1-\pi)\Phi_F(a)+\mathcal{I}(a>a^*)\pi\Phi_H(a). \label{Z.derivation}
\end{eqnarray}
This implies that the curve $P\{X \in \mathcal{B}(\mu_0, aV_0)\}$ has a change-point at $a^*$, since $\pi\Phi_H(a)$ can play a role only when $a>a^*$. For any $a\le a^*$, we have from Theorem~\ref{thm.consistency.contaminated} that $(\widehat\mu(a), \widehat V(a))$ provides an estimate of $(\mu_0,aV_0)$ and, hence,
\begin{eqnarray}
{\rm AR}(a)\approx P\{X \in \mathcal{B}(\mu_0, aV_0)\} \approx (1-\pi)\Phi_F(a),\quad \forall \quad a\le a^*. \label{AR.derivation}
\end{eqnarray}
It is reasonable to assume that $\psi(\cdot)$ eventually decreases to 0, in this situation $\Phi_F(a)$ and hence $\mathcal{C}_{\rm AR}$ is expected to be an increasing curve with decreasing slope when $a$ approaches $a^*$. When $a>a^*$, samples from $H$ start to play a role in $\widehat V(a)$, and (\ref{Z.derivation}) indicates that the increment size of $\mathcal{C}_{\rm AR}$ is expected to become larger than the case of $a\le a^*$. That is, an unusual large increment of $\mathcal{C}_{\rm AR}$ tends to appear due to the involvement of samples from $H$ that may deteriorate the performance of $\widehat V(a)$. From a conservative point of view, one should discard this portion of suspicious samples in order to achieve a reliable PCA result. We thus propose to identify $a^*$ via inspecting the increasing pattern of $\mathcal{C}_{\rm AR}$:
\begin{center}
{\it $\mathcal{C}_{\rm AR}$ should smoothly increase before $a^*$, but has unstable or relatively}\\
{\it large increments after $a^*$ due to the contamination effect of $H$.}
\end{center}

\noindent
The form of $H$ can be arbitrary. The effect of $H$, however, vanishes for small $a$, in this situation $\mathcal{C}_{\rm AR}$ is mainly driven by $F$ and, hence, can be more informatively used to identify $a^*$. This motivates the inspection of $\mathcal{C}_{\rm AR}$ from the left with small $a$ values.

To exemplify our idea, Figure~\ref{fig.ar_sim} reports $\mathcal{C}_{\rm AR}$ and the similarity measure $\rho$ from one simulation run under $(n,p,\pi)=(250,100,0.15)$ (see Section~\ref{sec.simulation} for detailed settings and the definition of $\rho$). For a rank-$k$ orthogonal matrix $\Gamma_k$ and its estimate $\widehat\Gamma_k$,
$\rho=1$ means that $\rmspan(\widehat\Gamma_k)=\rmspan(\Gamma_k)$, and $\rho=0$ means that $\widehat\Gamma_k \perp\Gamma_k$. We also fit a cubic spline curve to $\mathcal{C}_{\rm AR}$ to assist visual inspection.
The left panel shows the case of $c=3$ and the right panel shows the case of $c=2$, where a larger $c$ indicates a larger distance between $F$ and $H$. Summary results are given below.

\begin{enumerate}
\item
For the case of $c=3$, $\mathcal{C}_{\rm AR}$ smoothly increases to $72\%$ over $a\in(0.2,0.75)$, then becomes flat for $a\in(0.75,1.2)$, and starts to increase rapidly for $a\in (1.2,1.5)$. This conveys the following messages:
\begin{itemize}
\item
These $72\%$ sample points with $a$ values below $0.75$ are expected to belong to $F$ by the smoothly increasing pattern of $\mathcal{C}_{\rm AR}$ for $a\le 0.75$.

\item
There are very few sample points in the range corresponding to $a\in(0.75,1.2)$. Those remaining sample points (about  $28\%$) corresponding to $a>1.2$ are away from $F$ and are treated as outliers.
\end{itemize}
The above observations suggest that $a^*$ appears around $1.2$, and $\widehat V(a)$ with $a=1.2$ does not involve those $28\%$ suspicious sample points. The validity of the choice of $a=1.2$ can be seen from the fact that $\rho$ increases from $0.85$ to $0.92$ for $a\in(0.2,1.2)$ (which supports the claim that those $72\%$ of data mainly come from $F$), and that $\rho$ has a sudden decrease for $a\in(1.3,1.4)$ (which evidences the involvement of $H$ in the construction of $\widehat V(a)$ during this range of $a$).

\item
For the case of $c=2$, $\mathcal{C}_{\rm AR}$ increases smoothly to $67\%$ with a decreasing slope for $a\in(0.2,0.6)$, but starts to increase to $84\%$ quickly for $a\in(0.6,0.8)$. This conveys the following messages:
\begin{itemize}
\item
These $67\%$ sample points corresponding to $a\le 0.6$ are expected to belong to $F$, due to the small size of $a$ and the smoothly increasing pattern of $\mathcal{C}_{\rm AR}$ for $a\le 0.6$.

\item
Due to the different increasing patterns of $\mathcal{C}_{\rm AR}$ before and after $a=0.6$, the secondly included $33\%$ samples for $a>0.6$ are expected to come from $H$.
\end{itemize}
Unlike the case of $c=3$, where a flat region of $\mathcal{C}_{\rm AR}$ is observed for $a\in(0.75,1.2)$, one can only detect a sudden increase for the curve when $a>0.6$. The main reason is that when $c=2$, the supports of $F$ and $H$ heavily overlap, and there exists no range for $a$ with unchanged AR values. The sudden increase of $\mathcal{C}_{\rm AR}$, however, still conveys the message that for $a>0.6$, another bulk of heterogeneous samples start to play a role in the analysis, and this suggests that $a^*\approx 0.6$. The performance of the selection $a=0.6$ is again supported by a high $\rho$ value at $a=0.6$, and a sudden decrease of $\rho$ when $a>0.6$.
\end{enumerate}
It is worth emphasizing that large $\rho$ values can still be achieved at small $a$ values. This observation echoes Corollary~\ref{thm.estimating_equation.V.hard_threshold} that it suffices to recover $V_0$ from the samples in a neighborhood of $\mu_0$. This again suggests the tuning of SPPCA should be started from a small value of $a$.

We have shown that $\mathcal{C}_{\rm AR}$ can serve as a diagnostic tool to help determine $a^*$. To improve the procedure, we further propose a data-adaptive method to determine a candidate value of $a^*$. Firstly, it is reasonable to expect a decreasing trend for the slope of $\mathcal{C}_{\rm AR}$ when $a$ approaches $a^*$ from the left. When $a>a^*$, due to the involvement of $H$, we would expect an increasing trend for the slope of $\mathcal{C}_{\rm AR}$. These suggest that the slope of $\mathcal{C}_{\rm AR}$ tends to have a local minimum around $a^*$ (see also Figure~\ref{fig.ar_sim} for this phenomenon). Since $\mathcal{C}_{\rm AR}$ is expected to be a smoothly increasing curve for $a\le a^*$, we propose to approximate $a^*$ by {\it the first local minimum of the slope of $\mathcal{C}_{\rm AR}$} (when inspecting the curve from the left), denoted by $\widehat a^*$. Detailed implementation algorithm is summarized below.\\

\hrule  \vspace{0.25cm} {\noindent {\bf Algorithm 2.} (Implementation of choosing $\widehat a^*$) \vspace{0.25cm} \hrule
\begin{enumerate}
\item
Fit a cubic smoothing spline to $\mathcal{C}_{\rm AR}$ to obtain $\mathcal{C}_{\rm AR}^{(s)}$.

\item
Obtain from $\mathcal{C}_{\rm AR}^{(s)}$ the fitted AR values $\{{\rm AR}^{\rm (s)}(a):a\in\mathcal{A}\}$.

\item
Collect the points of local minimum
\[\mathcal{M}=\{a_j\in \mathcal{A} :{\rm AR}^{\rm (s)}(a_j)<{\rm AR}^{\rm (s)}(a_{j-1}),~{\rm AR}^{\rm (s)}(a_j)<{\rm AR}^{\rm (s)}(a_{j+1})\},\]
and output $\widehat a^*=\min \mathcal{M}$.
\end{enumerate}
\hrule  \vspace{0.7cm}

\noindent In Step-1, we obtain a smoothed version of $\mathcal{C}_{\rm AR}$ to avoid too many local minima due to data variation. The performance of $\widehat a^*$ is also reported in Figure~\ref{fig.ar_sim}. As expected, $\widehat a^*$ tends to associate with a relatively flat region of $\mathcal{C}_{\rm AR}$ (i.e., a region where the slope of $\mathcal{C}_{\rm AR}$ decreases), where an increasing trend can be further observed for the slope of $\mathcal{C}_{\rm AR}$ after $\widehat a^*$. A high $\rho$ value can also be observed at $\widehat a^*$, indicating the validity of $\widehat a^*$ as a suitable approximation of $a^*$. We remind the reader that the determination of $a^*$ requires a macro view of $\mathcal{C}_{\rm AR}$, and $\widehat a^*$ only provides a candidate value for $a^*$. It is still suggested to combine with $\mathcal{C}_{\rm AR}$ to have a better determination of $a^*$. We will further demonstrate the performance of $\widehat a^*$ via simulation studies in Section~\ref{sec.simulation}.

\begin{remark}\label{rmk.computation}
Given $\hat\mu(a)$ and $\widehat V(a)$, the computational cost for obtaining ${\rm AR}(a)$ in~(\ref{AR}) is to count how many observations belonging to the ball $\mathcal{B}(\hat\mu(a),\widehat V(a))$. This cost is ignorable. In fact, the proposed tuning procedure utilizes quantities $\hat\mu(a)$ and $\widehat V(a)$, which are already computed for SPPCA, and it requires few extra efforts in obtaining the $AR(a)$ values.
\end{remark}

\begin{remark}\label{rmk.AR}
To specify the set $\mathcal{A}$ in search for the optimal $a^*$, we let $\mathcal{C}_{\rm AR}$ run over a reasonable range. More specifically, for a given AR lower bound $\ell$, set $a_{\min}= \min\{a: {\rm AR}(a)\ge\ell\}$ and $a_{\max} =\min\{a:{\rm AR}(a_m) = 1\}$, i.e., the range of candidate AR values is $[\ell,1]$. The search range is a discretized set over $[a_{\min},a_{\max}]$ with equal space, say $\mathcal{A}=\{a_1,\dots,a_m\}$ with $a_1=a_{\min}$ and $a_m=a_{\max}$. In this article, we set $m=n/5$ and $\ell=20\%$.
\end{remark}

\section{Numerical Studies}\label{sec.simulation}

\subsection{Simulation settings}

The data $\{X_i\}_{i=1}^n$ are generated iid from the mixture distribution
\begin{eqnarray}
(1-\pi) t_\nu(0_p,V_0)+ \pi t_3(\mu_{\rm out},V_{\rm out}),
\label{sim.model}
\end{eqnarray}
where the main data distribution $t_\nu(0_p,V_0)$ is a $p$-variate $t$-distribution with degrees of freedom $\nu$, location parameter $0_p$ and shape parameter $V_0$, the contamination distribution is also a $t$-distribution $t_3(\mu_{\rm out},V_{\rm out})$ with location parameter $\mu_{\rm out}$ and shape parameter $V_{\rm out}$, and $\pi$ is the contamination proportion. Note that (\ref{sim.model}) violates the condition of central $a^*$-separability, since the support of $t_3(\mu_{\rm out},V_{\rm out})$ is the entire $\mathbb{R}^p$.
There are two sources of outliers in~(\ref{sim.model}) and the level of outlyingness is controlled by $(\nu,\pi)$:

\begin{itemize}
\item
Elliptical outliers: These are outliers due to the heavy-tails of $t_\nu(0_p,V_0)$, where a small $\nu$ leads to more outliers.
\item
Non-elliptical outliers: These are outliers from $t_3(\mu_{\rm o},V_{\rm out})$, where a large $\pi$ gives more outliers and $\pi=0$ gives no non-elliptical outliers.
\end{itemize}

In each simulation run, the parameters of interest $V_0=\sum_{j=1}^p\lambda_j\gamma_j\gamma_j^\top$ and the contamination distribution $t_3(\mu_{\rm out},V_{\rm out})$ are generated as follows:
\begin{itemize}
\item
For $\Gamma=[\gamma_1,\ldots,\gamma_p]$, we first generate a $p\times p$ matrix with entries iid from standard Gaussian, and then orthogonalize this matrix to get $\Gamma$.
\item
The signal eigenvalues
$\{\lambda_j\}_{j=1}^k$ are generated from the uniform distribution over $\left[2(1+(\frac{p}{n})^{\frac{1}{2}}), 10(1+(\frac{p}{n})^{\frac{1}{2}})\right]$.

\item
The noise eigenvalues $\{\lambda_j\}_{j=k+1}^p$ are generated from the uniform distribution over $[0,2]$.
\item
For the outlier distribution $t_3(\mu_{\rm out},V_{\rm out})$, we set $\mu_{\rm out}=c\sqrt{p}u$, where $u$ is sampled from the uniform distribution on the unit sphere ${\cal S}^{p-1}$, $V_{\rm out}$ is independently generated by the same mechanism as $V_0$, and $c$ controls the level of separation of the contamination distribution from the main data distribution.
\end{itemize}
This setting leads to the target of PCA to be the $k$-dimensional subspace $\rmspan(\Gamma_k)$, where $\Gamma_k$ consists of the first $k$ columns of $\Gamma$.

We implement SPPCA with the approximation $d(x,\mu,D_V)$, where $\widehat a^*$ from Algorithm~2 is determined by setting $\mathcal{A}$ to contain $50$ grid points on $[0.2p,3p]$ with equal space (denoted by SPPCA($\widehat a^*$)). For comparison, three existing robust PCA methods are considered, including Rocke-KSD, TME and ROBPCA. Maronna and Yohai (2017) have compared the performance of various robust scatter estimators, and found that the Rocke's estimator using KSD estimator (Pe\~{n}a and Prieto, 2007) as the initial value (denoted by Rocke-KSD) generally performs the best. Recall also that TME is the most robust shape estimator in the family of elliptical distributions. We thus implement Rocke-KSD (using the default code of Maronna and Yohai, 2017) and TME as potential competitors for SPPCA. The original TME requires to specify a robust estimator for $\mu$ and cannot be defined for $p>n$ due to the calculation of $d(x,\mu, V_{\rm T})$. For fair comparison, we thus implement TME using the approximation $d(x,\mu,D_{V_{\rm T}})$ discussed in Remark~\ref{rmk.fixed_point_algorithm} and set $\mu=\widehat \mu(\widehat a^*)$ from SPPCA($\widehat a^*$). The above-mentioned methods are based on the idea of conducting PCA on robust scatter estimators. Alternatively, we also implement ROBPCA (using the default code of Hubert, Rousseeuw, and Vanden Branden (2005)), a widely used robust PCA method based on the idea of projection pursuit, to compare with SPPCA.

For each robust PCA method that outputs a $k$-dimensional orthogonal basis $\widehat \Gamma_k$, its performance in estimating $\rmspan(\Gamma_k)$ is measured by the similarity $\rho=\frac{1}{k}\sum_{j=1}^ks_j$, where $s_j$ is the $j^{\rm th}$ singular value of $\widehat \Gamma_k^\top\Gamma_k$. Note that $\rho\in [0,1]$, $\rho=0$ indicates $\rmspan(\widehat \Gamma_k)\perp\rmspan(\Gamma_k)$, and $\rho=1$ indicates $\rmspan(\widehat \Gamma_k)=\rmspan(\Gamma_k)$. To demonstrate the potential ability of SPPCA, we also report $\rho$ of SPPCA with $a$ being selected to attain the maximal $\rho$ values over $\mathcal{A}$ (denoted by SPPCA(opt)). Note that SPPCA(opt) is not available in practice, and is used for the purpose of illustration only. Reported results for each method are the averaged $\rho$ values (over 100 replicates) under $n=250$ and all combinations of $\nu\in\{3,10\}$, $\pi\in\{0,0.15,0.3\}$, $c\in\{1,1.5,\ldots,4\}$, and $p\in\{100, 500\}$.

\subsection{Simulation results}

Figure~\ref{fig.sim1_p100} reports the simulation results of $p=100$. It can be seen that SPPCA($\widehat a^*$) is generally the best performer, followed by TME, ROBPCA, and Rocke-KSD. This coveys two messages. First, the high $\rho$ values of SPPCA(opt) indicate the potential ability of SPPCA to recover $\rmspan(\Gamma_k)$ in the presence of outliers, provided that $a$ is properly chosen. Second, SPPCA($\widehat a^*$) and SPPCA(opt) have similar $\rho$ values (except for the case of small $c$ when $\pi>0$), indicating the validity of using $\widehat a^*$ as a suitable approximation of $a^*$. Recall that SPPCA is robust to the specification of $\psi(\cdot)$ due to its semiparametric construction, and has vanishing IFs for outliers due to the hard-threshold exponential weight. The superiority of SPPCA($\widehat a^*$) can be more obviously seen in the presence of elliptical and/or non-elliptical outliers. In particular, SPPCA$(\widehat a^*)$ can achieve $\rho$ values around 0.9 for most settings of $(\nu,\pi)$ (except for the case of small $c$ when $\pi>0$), while $\rho$ of the rest methods (TME, ROBPCA, and Rocke-KSD) can have a moderate to large decay when $\nu$ decreases (i.e., more elliptical outliers) and/or $\pi$ increases (i.e., more non-elliptical outliers).

More comparisons for SPPCA and other methods are summarized below:
\begin{itemize}
\item
TME is the best performer when $\pi=0$. This is reasonable since TME is the most robust shape estimator under elliptical distributions and, hence, the performance of TME is not sensitive to the size of $\nu$ when $\pi=0$. When $\pi>0$, however, the ellipticity assumption of $X$ is violated, and the $\rho$ value of TME is found to decrease in $\pi$, even for the cases of large $c$ values (i.e., the non-elliptical outliers are far away from the main data), a consequence of ignoring the length of $X$ that contains useful information to identify outliers. On the other hand, SPPCA adopts the hard-threshold exponential weight that is able to utilize the information of the length of $X$. When $\pi>0$, it is thus reasonable to observe that SPPCA$(\widehat a^*)$ has $\rho$ value increases in $c$, and eventually outperforms TME when $c>1.5$.

\item
ROBPCA performs well for the simplest case $(\nu,\pi)=(10,0)$, but is sensitive to both types of outlyingness, where a low $\rho$ value is observed when $\nu$ decreases and/or $\pi$ increases. Similar to SPPCA, ROBPCA also uses AR as its tuning parameter. Without additional information to help tuning, however, ROBPCA (with the default AR value 0.75) is found to have poor performance in the presence of outliers. On the other hand, SPPCA determines $\widehat a^*$ via the increasing pattern of AR that is able to adapt to the underlying data characteristic. As a result, SPPCA$(\widehat a^*)$ outperforms ROBPCA, especially for the cases of $\nu=3$ and/or $\pi\ge0.15$.

\item
Rocke-KSD performs satisfactory for the simplest case $(\nu,\pi)=(10,0)$, but is sensitive to the presence of outliers and is the wort performer in all settings. Note that the tuning parameter of Rocke-KSD in its default implementation is determined to maintain the ARE with respect to the multivariate normal distribution. It is thus reasonable that Rocke-KSD performs well under the simplest case of $(\nu,\pi)=(10,0)$ (i.e., nearly multivariate normal). However, a large decay of $\rho$ can be observed when $\nu=3$ and/or $\pi\ge 0.15$ (i.e., violating the multivariate normal assumption). On the other hand, SPPCA determines $\widehat a^*$ without using the knowledge of the true distribution. The fact that SPPCA($\widehat a^*$) outperforms Rocke-KSD in all settings again indicates the adaptivity and robustness of SPPCA.
\end{itemize}

It should be emphasized that when $\pi>0$, SPPCA(opt) is still able to recover $\rmspan(\Gamma_k)$ (and achieves the highest $\rho$ values) even for the case of small $c$ (i.e., $F$ and $H$ are heavily overlapped). This observation echoes Corollary~\ref{thm.estimating_equation.V.hard_threshold} that it suffices for SPPCA to recover $V_0$ from the samples in $\mathcal{B}(\mu_0,aV_0)$ with a small $a$. However, we can also observe that SPPCA$(\widehat a^*)$ cannot perform comparable with SPPCA(opt) in this situation. Indeed, a small $c$ indicates a large overlapping for the main and outlier distributions, which makes $\mathcal{C}_{\rm AR}$ not having a clear change-point. As a result, SPPCA$(\widehat a^*)$ cannot approach SPPCA(opt) due to the poor approximation of $\widehat a^*$ to $a^*$. Nevertheless, for the case of small $c$, we can still observe SPPCA$(\widehat a^*)$ has comparable performance with TME, and outperforms ROBPCA and Rocke-KSD even with a less accurate approximation $\widehat a^*$.

The simulation results of $p=500$ are reported in Figure~\ref{fig.sim1_p500}. Rocke-KSD  cannot be implemented when $p>n$ and is omitted. Generally speaking, we can draw similar conclusion as the case of $p=100$ that SPPCA$(\widehat a^*)$ is the best performer, followed by TME and ROBPCA. We still observe similar performances for SPPCA$(\widehat a^*)$ and SPPCA(opt), and the stable $\rho$ values around 0.8 for SPPCA$(\widehat a^*)$ (except for the case of small $c$ when $\pi>0$). TME is again found to be the best performer when $\pi=0$, but can miss certain directions of $\rmspan(\Gamma_k)$ when $\pi>0$. Recall that for the sake of robustness, we use the approximation $d(x,\mu,D_V)$ to implement SPPCA (and TME). The success of using $d(x,\mu,D_V)$ is confirmed from the high $\rho$ values of SPPCA$(\widehat a^*)$, and by noting that the $\rho$ values of SPPCA$(\widehat a^*)$ under $p=500$ is slightly smaller than the case of $p=100$. ROBPCA, however, is found to behave poorly, and has a huge decay for $\rho$ as compared to the case of $p=100$, indicating the instability of ROBPCA in the high-dimensional case. In summary, our simulation results reveal the robustness and adaptivity of SPPCA$(\widehat a^*)$ to the presence of outliers without sacrificing much its estimation efficiency, and is able to perform well even for the case of $p>n$.

\section{Data Analysis}\label{sec.data}

We analyze the data set of Wu {\it et al.} (2011) that was later analyzed in Zheng, Lv and Lin (2021). The aim of this data analysis is to investigate the association between BMI (denoted by $Y$) and habitual diet effect in the human gut microbiome. The data set consists of $n = 98$ subjects, each with $p=301$ measurements (214 for nutrient intake and 87 for gut microbiome composition). The scales of nutrient intake and gut microbiome composition differ a lot, and we preprocess the data by component-wise standardization (denoted by $X$) to enter the analysis.

Since ${\rm AR}(a)=0$ for $a\le 0.5p$ and ${\rm AR}(a)\approx 1$ for $a>9p$, we set $\mathcal{A}$ to contain 20 equally-spaced grid points on $[0.5p, 9p]$. The resulting $\mathcal{C}_{\rm AR}$ and $\widehat a^*=2.74p$ from Algorithm~2 are reported in Figure~\ref{fig.data}(a), which gives AR$(\widehat a^*)=0.8163$. The slope of $\mathcal{C}_{\rm AR}$ shows a decreasing trend when $a$ increases to $\widehat a^*$, and then starts to increase when $a>\widehat a^*$. This pattern indicates that those $18.37\%$ sample points with $a>\widehat a^*$ likely come from a heterogeneous distribution. We thus proceed to conduct SPPCA at the scale $\widehat a^*$. Let $\widehat\Gamma_k$ be the leading $k$ eigenvectors of $\widehat V(\widehat a^*)$. Figure~\ref{fig.data}(b) displays the 3-dimensional scatter plot of the principal component scores
$\{\widehat\Gamma_3^\top X_i\}_{i=1}^{98}$, wherein 18 sample points receiving zero weights are circled. These circled sample points tend to be away from the main data cloud. It indicates that the population could be a mixture of two distributions with different locations.
%{\red\sout{This observation again supports the exclusion of these 18 suspicious sample %points from the PCA  to achieve a better dimension reduction result for the rest $80$ %sample points.}}

To further demonstrate the performance of $\widehat V(\widehat a^*)$, we fit a linear regression model to $\{(Y_i,\widehat\Gamma_k^\top X_i)\}_{i=1}^{98}$, and report the adjusted-$R^2$ values for $k\in\{1,2,\ldots,40\}$ in Figure~\ref{fig.data}(c). Results from TME, ROBPCA, and the usual PCA  to construct $\widehat\Gamma_k$ are also reported for comparisons. SPPCA achieves its maximum adjusted-$R^2$ value $0.2974$ at $k=28$, followed by $0.2874$ of TME at $k=28$, $0.2377$ of PCA at $k=34$, and $0.1941$ of ROBPCA at $k=27$. Detailed comparisons for SPPCA and other methods are listed below:

\begin{itemize}
\item
SPPCA outperforms TME for most of the $k$ values. Recall that the scatter plot in Figure~\ref{fig.data}(b) shows the violation of the ellipticity for the data. It is thus reasonable that TME has a worse fitting result, since TME is not able to mitigate the effects of non-elliptical outliers. On the other hand, SPPCA assigns zero weights for outliers, and can achieve a higher adjusted-$R^2$ value than TME.

\item
ROBPCA fails to extract informative directions for dimension reduction. This again echoes our findings from the simulation studies that ROBPCA is very sensitive to the presence of outliers. As a result, ROBPCA behaves similarly with the usual PCA, and its optimal adjusted-$R^2$ is even smaller than the usual PCA's. An issue to implement ROBPCA is the selection of AR (default 0.75). The analysis results of ROBPCA under different AR values are reported in Figure~\ref{fig.data}(d). One can see that ROBPCA performs the best when AR equals $0.5$, which achieves its maximum adjusted-$R^2$ value 0.2662 at $k=25$. This conveys two messages. First, the default AR value 0.75 for ROBPCA tends to be subjective and cannot adapt to the underlying data characteristics. That is, ROBPCA demands an informative tuning procedure to ensure its performance. Second, SPPCA with the data-adaptive $\widehat a^*$ outperforms ROBPCA regardless the choice of its AR value.
\end{itemize}

The SPPCA algorithm outputs 18 observations with zero weights. These 18 data points  (circled in Figure~\ref{fig.data}(b)) can be identified as outliers. The adjusted-$R^2$ values from fitting a linear regression model to $(Y,\widehat\Gamma_k^\top X)$ based on the rest 80 active sample points (denoted by SPPCA+) are also reported in Figure~\ref{fig.data}(a). It can be seen that SPPCA+ outperforms SPPCA for every $k$, and achieves the maximum adjusted-$R^2$ 0.3172 at $k=28$. This observation indicates the existence of heterogeneity regarding the association between $X$ and $Y$. By excluding the 18 suspicious sample points, we can better explain $Y$ by $\widehat\Gamma_k^\top X$.

\section{Discussions}\label{sec.discussion}

In this article, we propose a robust SPPCA method. By utilizing the semiparametric theory, we develop a class of estimating equations for $(\mu_0,V_0)$. These estimating equations do not depend on the radial function $\psi(\cdot)$. By further equipped with the hard-threshold exponential weight function $w(\cdot)$, SPPCA is shown to be robust to the presence of both elliptical and non-elliptical outliers. We also develop an AR-based tuning procedure to facilitate the implementation of SPPCA. Conducting SPPCA based on $\widehat V(\widehat a^*)$ is almost free of tuning parameters, except for the choice of the candidate set $\mathcal{A}$ to construct $\mathcal{C}_{\rm AR}$.
We observe that the performance of SPPCA is quite stable to the choice of $\mathcal{A}$.
%This also makes SPPCA significantly different from existing methods, considering the fact that there generally lacks informative and scalable method to tuning a robust shape estimator indexed by AR.

A data-adaptive SPPCA is proposed to perform PCA based on $\widehat V(\widehat a^*)$. As demonstrated in our numerical studies, while SPPCA($\widehat a^*$) outperforms existing methods in the presence of outliers, we also detect a gap between SPPCA($\widehat a^*$) and SPPCA(opt) when the supports of $F$ and $H$ are heavily overlapped. This indicates that the less accurate performance of SPPCA($\widehat a^*$) is mainly due to a poor selection of $\widehat a^*$. It is thus of interest to further investigate the estimation of $a^* $ under heavy overlapping of $F$ and $H$ to improve SPPCA in a future study.

For the case of $p>n$, we conveniently used $D_V^{-1}$ to replace $V^{-1}$ in $d(x,\mu,V)$ (see Remark~\ref{rmk.fixed_point_algorithm}) for both SPPCA and TME. There are regularized versions of TME
(Chen, Wiesel, and Hero, 2011; Sun, Babu, and Palomar, 2014; and Ollila and Tyler, 2014). Following the same idea, we can also consider a regularized SPPCA for the case of $p>n$ by adopting the regularized estimating equation of $V$:
\begin{eqnarray}
V = \frac{1}{1+\tau}\frac{p\int w\left(d(x,\mu,V)\right)(x-u)^{\otimes 2} d F}
{\int w\left(d(x,\mu,V)\right)d(x,\mu,V) d F}+\frac{\tau}{1+\tau}I_p, \label{semi-PCA.regularized}
\end{eqnarray}
where $\tau\ge 0$ is a regularization parameter, and (\ref{semi-PCA.regularized}) reduces to (\ref{estimating_equation.V}) when $\tau=0$. A practical issue of (\ref{semi-PCA.regularized}) is the determination of $\tau$. For the case of regularized TME, Ollila and Tyler (2014) proposed to estimate $\tau$ via minimizing the mean squared error. This criterion, however, does not adapt to $(\pi,H)$ and may produce unreliable analysis results. In view of this point, we used $d(x,\mu,D_V)$ to approximate $d(x,\mu,V)$ in this work, owing to the fact that this strategy does not involve extra tuning parameters. However, (\ref{semi-PCA.regularized}) still possesses its own merit by preserving the covariance structure when calculating $d(x,\mu,V)$. It is of interest to investigate the statistical inference procedure of the regularized SPPCA based on (\ref{semi-PCA.regularized}) in a future study.

\section*{References}
\begin{description}
\item
Chen, Y., Wiesel, A., and Hero, A. O. (2011). Robust shrinkage estimation of high-dimensional covariance matrices. {\it IEEE Transactions on Signal Processing}, 59(9), 4097-4107.

\item
Chen, M., Gao, C., and Ren, Z. (2018). Robust covariance and scatter matrix estimation under Huber’s contamination model. {\it The Annals of Statistics}, 46(5), 1932-1960.

\item
Croux, C., Garc­\'{i}a-Escudero, L. A., Gordaliza, A., Ruwet, C., and Mart\'{i}n, R. S. (2017). Robust principal component analysis based on trimming around affine subspaces. {\it Statistica Sinica}, 27, 1437-1459.

\item
Davies, P. L. (1987). Asymptotic behaviour of S-estimates of multivariate location parameters and dispersion matrices. {\it Annals of Statistics}, 15(3), 1269-1292.

%\item
%Frahm, G., and Jaekel, U. (2010). A generalization of Tyler's M-estimators to the case of %incomplete data. {\it Computational Statistics \& Data Analysis}, 54(2), 374-393.

\item
Fortunati, S., Renaux, A., and Pascal, F. (2020). Robust semiparametric efficient estimators in complex elliptically symmetric distributions. {\it IEEE Transactions on Signal Processing}, 68, 5003-5015.

\item
Fortunati, S., Renaux, A., and Pascal, F. (2022). Joint Estimation of Location and Scatter in Complex Elliptically Symmetric Distributions. {\it Journal of Signal Processing Systems}, 94(2), 133-146.

\item
Gnanadesikan, R. and Kettenring, J. R. (1972). Robust estimates, residuals, and outlier detection with multiresponse data. {\it Biometrics}, 28, 81-124.

\item
Goes, J., Lerman, G., and Nadler, B. (2020). Robust sparse covariance estimation by thresholding Tyler’s M-estimator. {\it The Annals of Statistics}, 48(1), 86-110.

%\item
%Hettmansperger, T. P., and Randles, R. H. (2002). A practical affine equivariant multivariate %median. {\it Biometrika}, 89, 851-860.

\item
Hallin, M., Oja, H., and Paindaveine, D. (2006). Semiparametrically efficient rank-based inference for shape. II. Optimal R-estimation of shape. {\it The Annals of Statistics}, 34(6), 2757-2789.

\item
Han, F. and Liu, H. (2018). ECA: High-dimensional elliptical component analysis in non-Gaussian distributions. {\it Journal of the American Statistical Association}, 113(521), 252-268.

\item
Hubert, M., Rousseeuw, P. J., and Vanden Branden, K. (2005). ROBPCA: a new approach to robust principal component analysis. {\it Technometrics}, 47(1), 64-79.

%\item
%Lopuha\"a, H. P. (1989). On the relation between S-estimators and M-estimators of multivariate location and %covariance. {\it Annals of Statistics}, 17(4), 1662-1683.

\item
Lopuha\"a, H. P. (1991). Multivariate $\tau$-estimators for location and scatter. {\it Canadian Journal of Statistics}, 19(3), 307-321.

\item
Hallin, M., Paindaveine, D., and Verdebout, T. (2014). Efficient R-estimation of principal and common principal components. {\it Journal of the American Statistical Association}, 109(507), 1071-1083.

\item
Marden, J. I. (1999). Some robust estimates of principal components. {\it Statistics \& probability letters}, 43(4), 349-359.

\item
Maronna, R. A. (1976). Robust M-estimators of multivariate location and scatter. {\it Annals of Statistics}, 4(1), 51-67.

\item
Maronna, R. A. and Zamar, R. H. (2002). Robust estimates of location and dispersion for high-dimensional datasets. {\it Technometrics}, 44(4), 307-317.

\item
Maronna, R. A., and Yohai, V. J. (2017). Robust and efficient estimation of multivariate scatter and location. {\it Computational Statistics \& Data Analysis}, 109, 64-75.

\item
Ollila, E. and Tyler, D. E. (2014). Regularized M-estimators of scatter matrix. {\it IEEE Transactions on Signal Processing}, 62(22), 6059-6070.

\item
Ollila, E., Palomar, D. P., and Pascal, F. (2020). Shrinking the eigenvalues of M-estimators of covariance matrix. {\it IEEE Transactions on Signal Processing}, 69, 256-269.

%\item
%Paindaveine, D., and Van Bever, G. (2018). Halfspace depths for scatter, concentration %and shape matrices. {\it The Annals of Statistics}, 46(6B), 3276-3307.

\item
Paindaveine, D., and Van Bever, G. (2019). Tyler shape depth. {\it Biometrika}, 106(4), 913-927.

\item
Pe\~{n}a, D., and Prieto, F. J. (2007). Combining random and specific directions for outlier detection and robust estimation in high-dimensional multivariate data. {\it Journal of Computational and Graphical Statistics}, 16(1), 228-254.

\item
Rocke, D. M. (1996). Robustness properties of S-estimators of multivariate location and shape in high dimension. {\it Annals of Statistics}, 24(3), 1327-1345.

\item
Rousseeuw, P. J. (1985). Multivariate estimation with high breakdown point. {\it Mathematical Statistics and Applications}, 8(37), 283-297.

\item

Salibi{\'a}n-Barrera, M., Van Aelst, S. and Willems, G. (2006). Principal components analysis based on multivariate MM estimators with fast and robust bootstrap. {\it Journal of the American Statistical Association}, 2006, 101(475), 1198-1211.

\item
Sun, Y., Babu, P., and Palomar, D. P. (2014). Regularized Tyler's scatter estimator: Existence, uniqueness, and algorithms. {\it IEEE Transactions on Signal Processing}, 62(19), 5143-5156.

\item
Tsiatis, A. (2007). {\it Semiparametric Theory and Missing Data}. Springer Science \& Business Media.

\item
Tatsuoka, K. S. and Tyler, D. E. (2000). On the uniqueness of S-functionals and M-functionals under nonelliptical distributions. {\it Annals of Statistics}, 28(4), 1219-1243.

\item
Tyler, D. E. (1987). A distribution-free M-estimator of multivariate scatter. {\it Annals of Statistics}, 15(1), 234-251.

%\item
%Tyler, D. E. (1987b). Statistical analysis for the angular central Gaussian distribution on %the sphere. {\it Biometrika}, 74(3), 579-589.

\item
Wu, G. D., Chen, J., Hoffmann, C., Bittinger, K., Chen, Y. Y., Keilbaugh, S. A., Bewtra, M., Knights, D., Walters, W. A., Knight, R.,
Sinha, R., Gilroy, E., Gupta, K., Baldassano, R., Nessel, L., Li, H., Bushman, F. D., and Lewis, J. D. (2011).
Linking long-term dietary patterns with gut microbial enterotypes. {\it Science}, 334, 105-108.

\item
Zheng, Z., Lv, J. and Lin, W. (2021). Nonsparse learning with latent variables. {\it Operations Research}, 69(1), 346-359.
\end{description}

\newpage

\begin{figure}[!ht]
\centering
\includegraphics[width=3.1in]{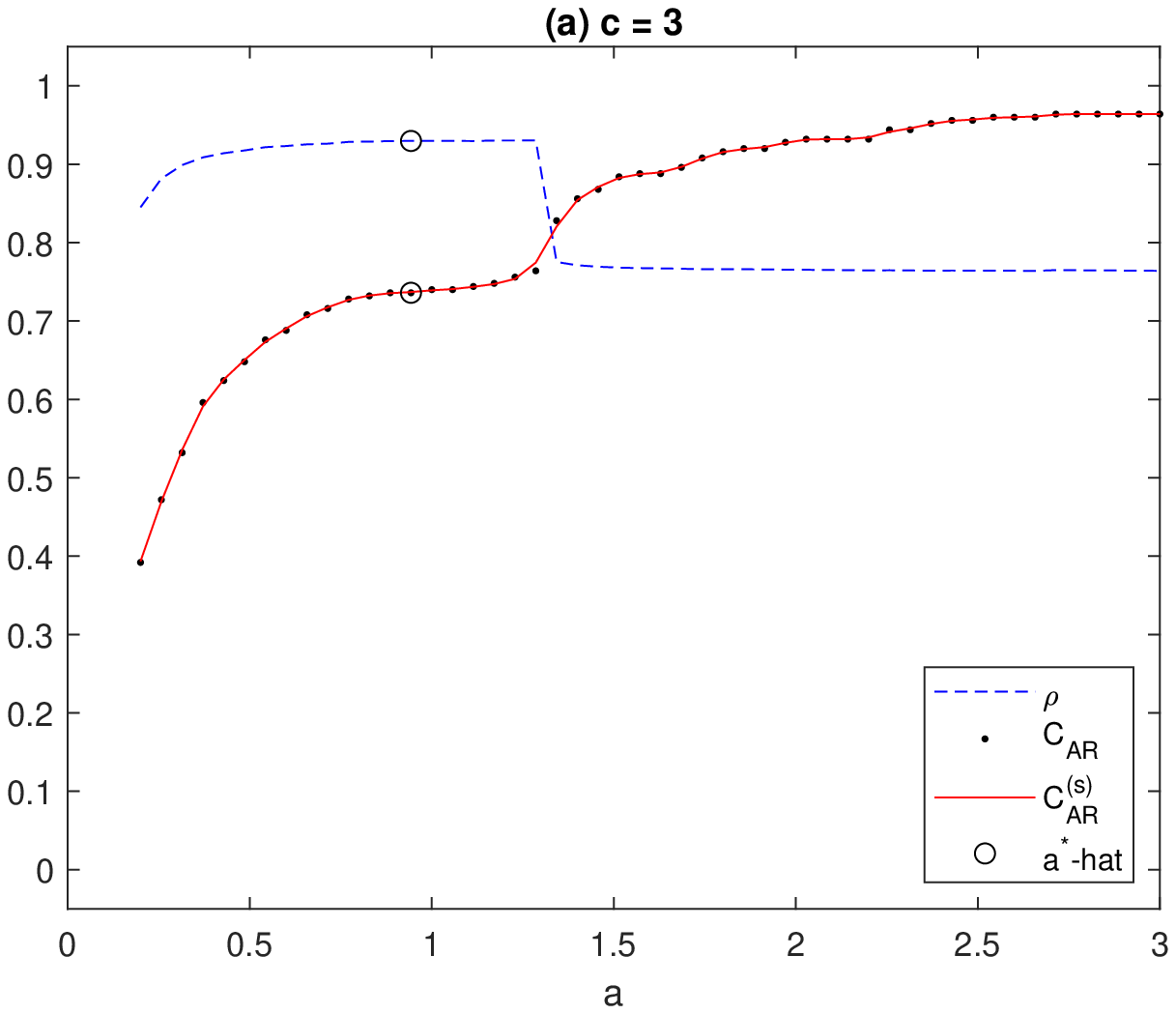}
\includegraphics[width=3.1in]{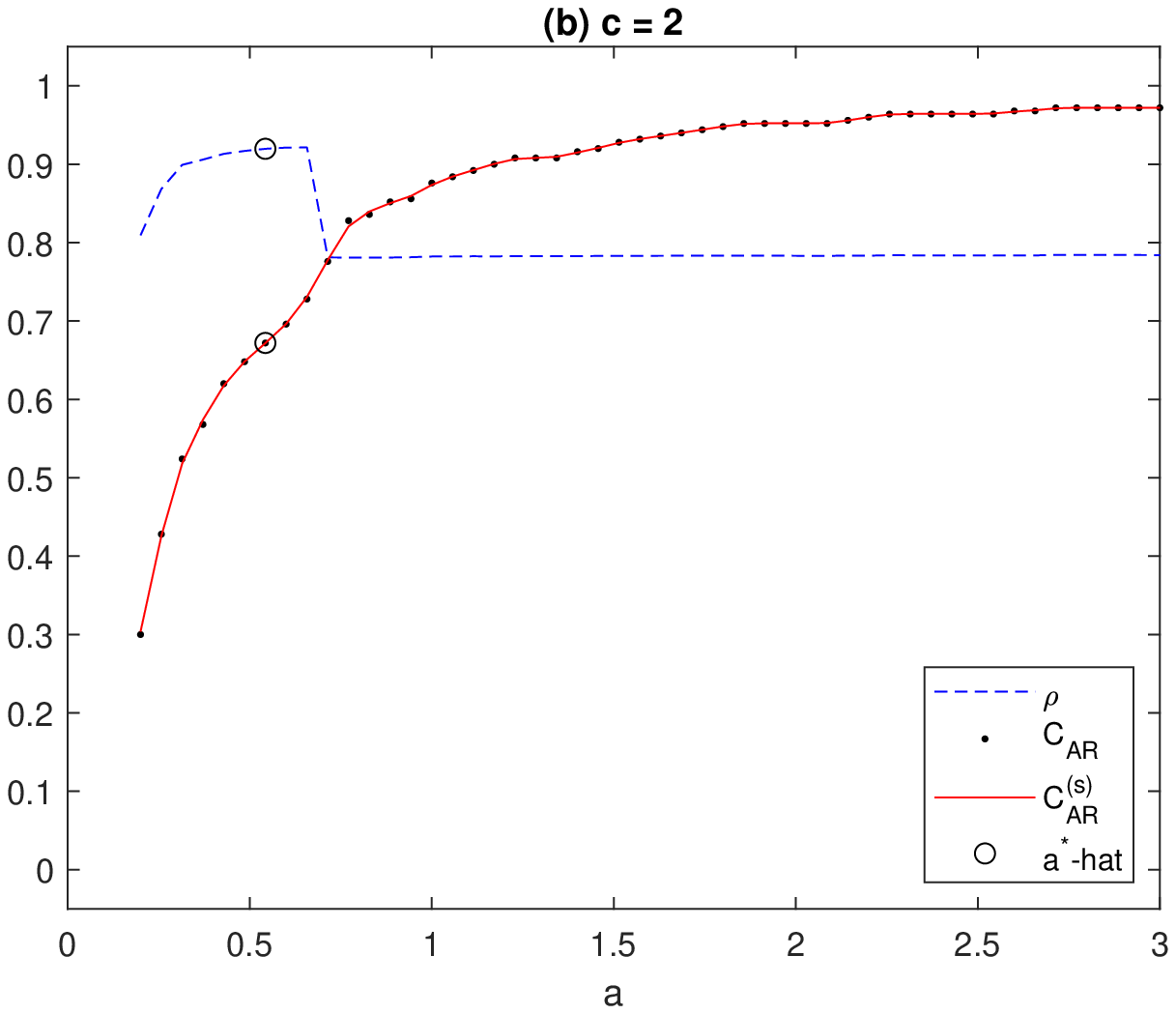}
\caption{The AR curve $\mathcal{C}_{\rm AR}$ (black dots) and  its cubic smoothing spline fit $\mathcal{C}_{\rm AR}^{(s)}$ (red real curve) from one simulation run under $(n,p,\pi)=(250,200,0.15)$ and $c=2,3$. The corresponding similarity values $\rho$ at different scale size $a$ are also reported (blue dash curve). The circle on $\mathcal{C}_{\rm AR}$ indicates the position of the data-adaptive $\widehat a^*$ from Algorithm~2, and the corresponding $\rho$ value of $\widehat V(\widehat a^*)$ is also circled. (a) The case of $c=3$, where the supports of $F$ and $H$ are almost completely separated. (b) The case of $c=2$, where the supports of $F$ and $H$ are heavily overlapped.}\label{fig.ar_sim}
\end{figure}

\newpage

\begin{figure}[!ht]
\hspace{-1.7cm}
\includegraphics[width=7.6in]{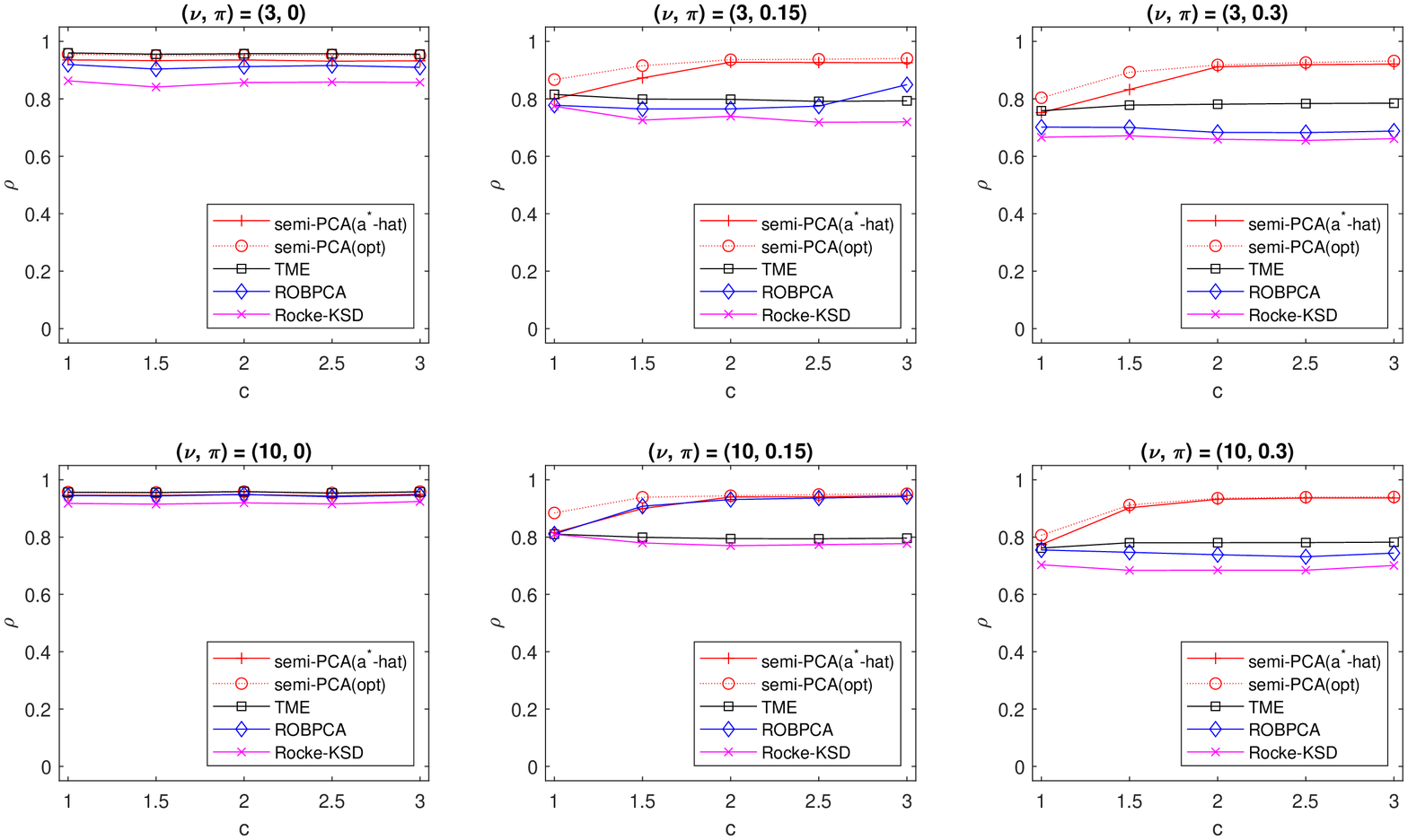}
\caption{The similarity values $\rho$ for different methods, SPPCA($\widehat a^*$), SPPCA(opt), TME, ROBPCA, and Rocke-KSD, over different $c$ values under $(n,p)=(250,100)$ and different combinations of $\nu\in\{3,10\}$ and $\pi\in\{0,0.15,0.3\}$. Note that a smaller $\nu$ gives more elliptical outliers from $t_\nu(0,V_0)$, a larger $\pi$ gives more non-elliptical outliers from $t_3(\mu_{\rm out},V_{\rm out})$, and a larger $c$,  where $\mu_{\rm out}=c\sqrt{p}$, indicates a larger distance between $t_\nu(0,V_0)$ and $t_3(\mu_{\rm out},V_{\rm out})$.}\label{fig.sim1_p100}
\end{figure}

\newpage

\begin{figure}[!ht]
\hspace{-1.7cm}
\includegraphics[width=7.6in]{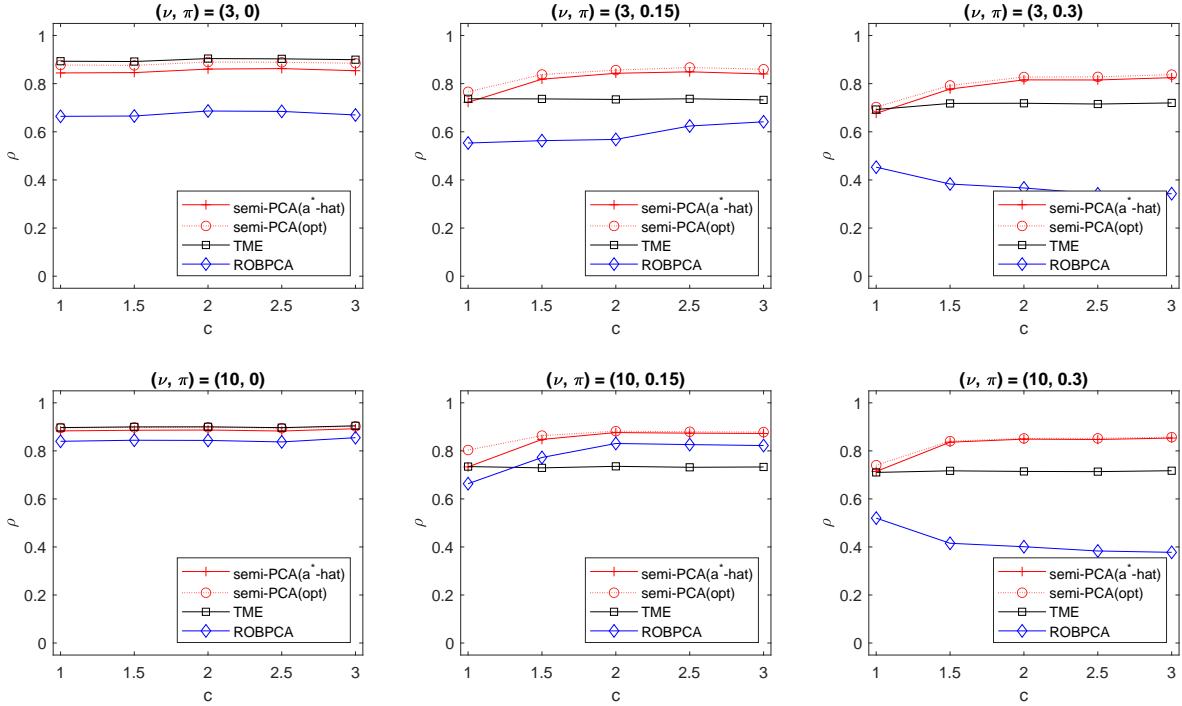}
\caption{The similarity values $\rho$ for different methods, SPPCA($\widehat a^*$), SPPCA(opt), TME, and ROBPCA, over different $c$ values under $(n,p)=(250,500)$ and different combinations of $\nu\in\{3,10\}$ and $\pi\in\{0,0.15,0.3\}$. Note that a small $\nu$ gives more elliptical outliers from $t_\nu(0,V_0)$, a larger $\pi$ gives more non-elliptical outliers from $t_3(\mu_{\rm out},V_{\rm out})$, and a larger $c$, where $\mu_{\rm out}=c\sqrt{p}$, indicates a larger distance between $t_\nu(0,V_0)$ and $t_3(\mu_{\rm out},V_{\rm out})$.}\label{fig.sim1_p500}
\end{figure}

\newpage
\pagenumbering{gobble}
\begin{figure}[!ht]
\hspace{-1.3cm}
\includegraphics[width=7.3in,height=3.2in]{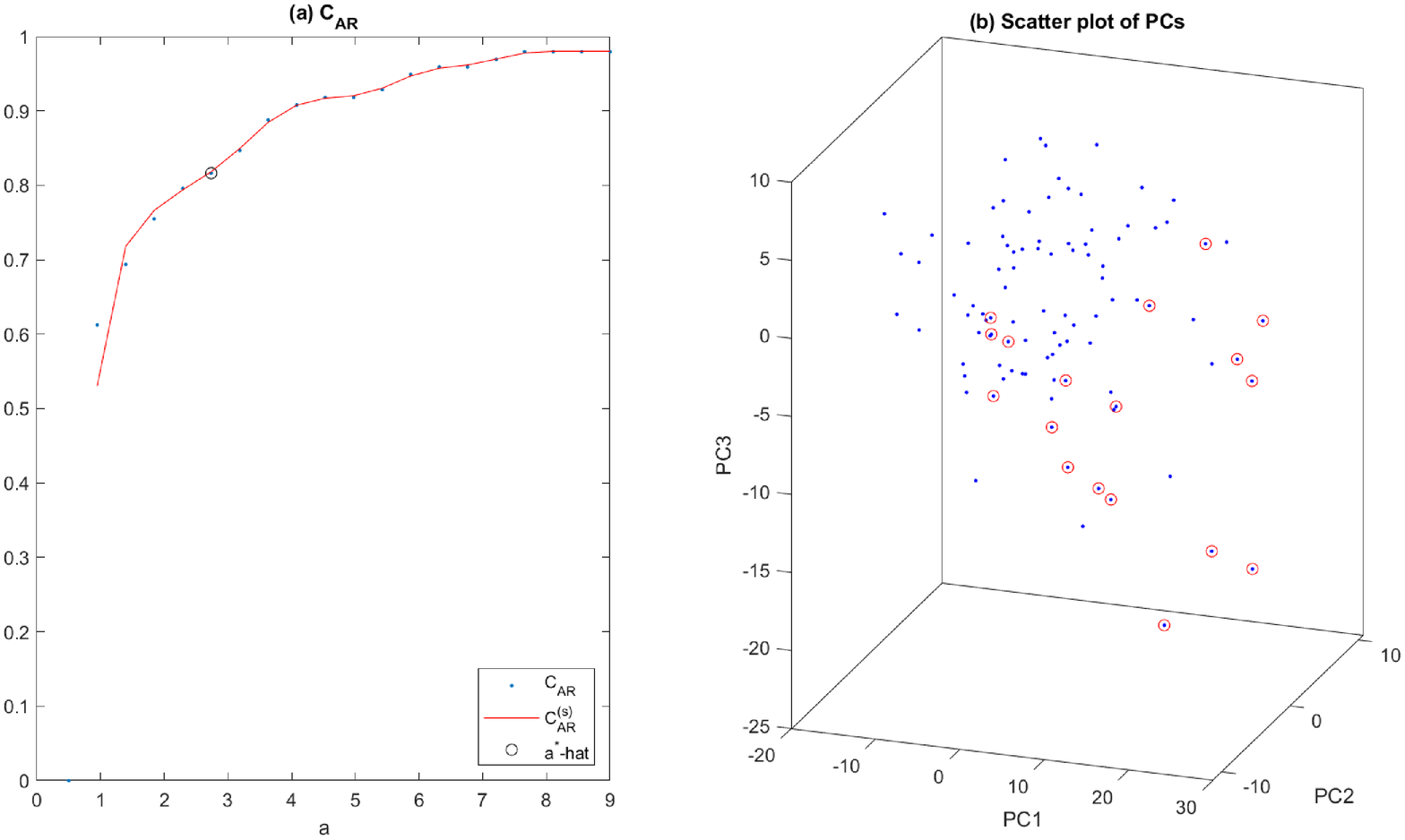}

\hspace{-1.3cm}
\includegraphics[width=7.3in,height=3.2in]{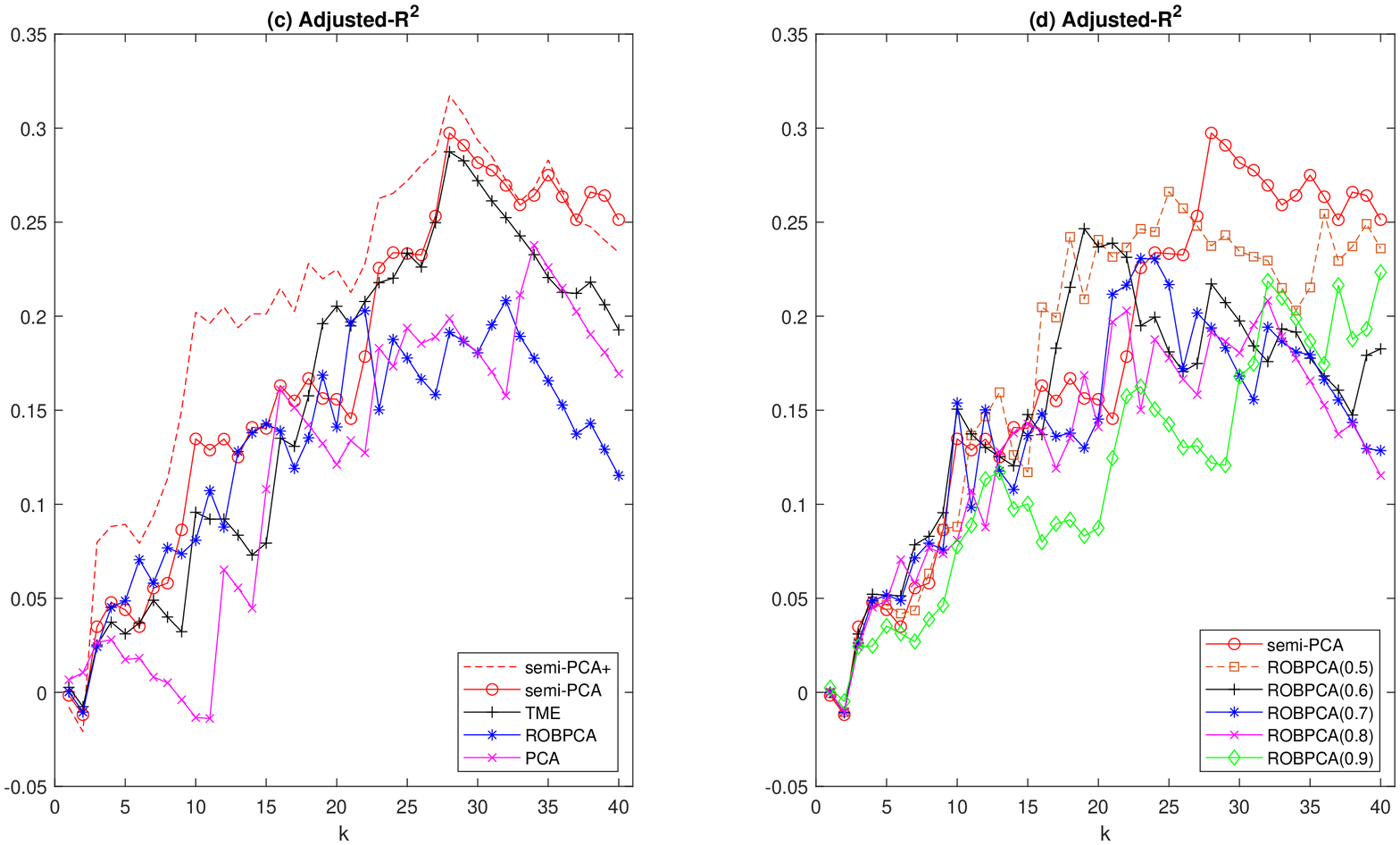}
\vspace{-1.5cm}
\caption{(a) The AR curve $\mathcal{C}_{\rm AR}$ (blue dots) and its cubic smoothing spline fit $\mathcal{C}_{\rm AR}^{(s)}$ (red real curve). The circle indicates the position of the data-adaptive $\widehat a^*$ from Algorithm~2. (b) The 3-dimensional scatter plot of the principal component scores $\{\widehat\Gamma_3^\top X_i\}_{i=1}^{98}$ from SPPCA at $\widehat a^*$, where the samples receiving zero weights are circled. (c) The adjusted-$R^2$ from fitting a linear regression model on $\{(Y_i, \widehat\Gamma_k^\top X_i)\}_{i=1}^{98}$ for $k\in\{1,2,\ldots,40\}$, where $\widehat\Gamma_k$ are obtained by SPPCA, TME, ROBPCA and the usual PCA. The adjusted-$R^2$ for a linear fitting using 80 data points (18 outlying points are excluded from linear fit) is denoted by SPPCA+. (d) The adjusted-$R^2$ from fitting a linear regression model on $\{(Y_i, \widehat\Gamma_k^\top X_i)\}_{i=1}^{98}$ for $k\in\{1,2,\ldots,40\}$, where $\widehat\Gamma_k$ are obtained from either SPPCA or ROBPCA using different AR values in $\{0.5,0.6,\ldots,0.9\}$.}\label{fig.data}
\end{figure}

\end{document}